\documentclass[floats,floatfix,amssymb,prd,twocolumn,nofootinbib,superscriptaddress]{revtex4-1}
\usepackage{amssymb,amsmath,verbatim,mathtools,needspace,enumitem,etoolbox,physics,microtype,ulem}
\usepackage[utf8]{inputenc}
\usepackage{graphicx}

\usepackage{booktabs}
\usepackage{xspace}
\usepackage{subcaption}
\usepackage{csvsimple}
\usepackage{pifont}
\usepackage{appendix}
\usepackage{multirow}
\usepackage[dvipsnames, usenames]{xcolor}
\usepackage{pgfplotstable}
\usepackage{float}
\usepackage[unicode, colorlinks=true, linkcolor=linkcolor, citecolor=linkcolor, filecolor=linkcolor,urlcolor=linkcolor, pdfusetitle]{hyperref}
\pagecolor{white}
\usepackage{aas_macros}
\usepackage{multirow}
\usepackage{sub caption}
\usepackage{cleveref}
\usepackage{natbib,tabularx}
\usepackage{booktabs,subcaption}%

\usepackage{gensymb} %
\usepackage{amsbsy} %

\definecolor{linkcolor}{rgb}{0.0,0.3,0.5}
\definecolor{urlcolor}{rgb}{0.27,0.55,0.}
\definecolor{funcolor}{rgb}{0.65, 0.16, 0.16}

\newcommand{\hubble}{\ensuremath{H_0}\xspace}
\newcommand{\dl}{\ensuremath{D_L}\xspace}

\newcolumntype{b}{>{\hsize=0.6\hsize\centering\arraybackslash}X}
\newcolumntype{Y}{>{\hsize=1.1\hsize\centering\arraybackslash}\centering X}

\newcommand{\mchirp}{\ensuremath{\mathcal{M}}\xspace}

\newcommand{\rmod}{\ensuremath{R_\mathrm{model}}\xspace}
\newcommand{\rtrue}{\ensuremath{R_\mathrm{true}}\xspace}
\newcommand{\etaOut}{\ensuremath{\{\eta^\mathrm{out}\}}\xspace}
\newcommand{\etaTyp}{\ensuremath{\{\eta^\mathrm{typ}\}}\xspace}
\newcommand{\miscal}{\ensuremath{\eta^\mathrm{mis}}\xspace}

\newcommand{\rmodIfo}{\ensuremath{R_\mathrm{model,IFO}}\xspace}
\newcommand{\rtrueIfo}{\ensuremath{R_\mathrm{true,IFO}}\xspace}

\newcommand{\ed}{\ensuremath{\boldsymbol{e}_{\mathrm{IFO}}}\xspace}
\newcommand{\boldd}{\ensuremath{\boldsymbol {d}}\xspace}
\newcommand{\ifod}{\ensuremath{\boldsymbol {d}_\mathrm{IFO}}\xspace}
\newcommand{\boldth}{\ensuremath{\boldsymbol {\theta}}\xspace}
\newcommand{\dmis}{\ensuremath{{\boldsymbol {d}}^{\rm mis}}\xspace}

\newcommand{\llhd}{\ensuremath{\mathcal{L}\xspace}}

\newcommand{\beqn}{\begin{eqnarray}}
\newcommand{\enqn}{\end{eqnarray}}
\newcommand{\beq}{\begin{equation}}
\newcommand{\eeq}{\end{equation}}

\newcommand{\LIGOlabMIT}{\affiliation{LIGO Laboratory, Massachusetts Institute of Technology, 185 Albany St, Cambridge, MA 02139, USA}}
\newcommand{\MKI}{\affiliation{Department of Physics and Kavli Institute for Astrophysics and Space Research, Massachusetts Institute of Technology, 77 Massachusetts Ave, Cambridge, MA 02139, USA}}

\newcommand{\UNLVphysics}{\affiliation{Department of Physics and Astronomy, University of Nevada, Las Vegas, 4505 South Maryland Parkway, Las Vegas, NV 89154, USA}}
\newcommand{\NCfA}{\affiliation{Nevada Center for Astrophysics, University of Nevada, Las Vegas, NV 89154, USA}}
\newcommand\UT{\affiliation{Department of Physics, The University of Texas at Austin, 2515 Speedway, Austin, TX 78712, USA}}

\pgfplotsset{compat=1.16}
\begin{document}
\title{Impact of calibration uncertainties on Hubble constant measurements from gravitational-wave sources}

\author{Yiwen Huang} 
\email{ywh@mit.edu, evahuangyiwen@gmail.com}
\LIGOlabMIT \MKI

\author{Hsin-Yu Chen} 
\LIGOlabMIT \MKI \UT

\author{Carl-Johan Haster} 
\LIGOlabMIT \MKI \UNLVphysics \NCfA

\author{Ling Sun}
\affiliation{OzGrav-ANU, Centre for Gravitational Astrophysics, College of Science, The Australian National University, ACT 2601, Australia}

\author{Salvatore Vitale}
\email{salvatore.vitale@ligo.mit.edu}
\LIGOlabMIT \MKI

\author{Jeffrey S. Kissel} 
\affiliation{LIGO Hanford Observatory, Richland, Washington 99352, USA}

\date{\today}

\begin{abstract}
Gravitational-wave (GW) detections of electromagnetically bright compact binary coalescences can provide an independent measurement of the Hubble constant \hubble. In order to obtain a measurement that could help arbitrate the existing tension on \hubble, one needs to fully understand any source of systematic biases for this approach. 
In this study, we aim to understand the impact of instrumental calibration errors (CEs) on the measurements of the luminosity distance, \dl, and the inferred \hubble value. We simulate binary neutron star mergers (BNSs), as detected by a network of Advanced LIGO and Advanced Virgo interferometers at their design sensitivity. We artificially add CEs equal to exceptionally large values experienced in LIGO-Virgo's third observing run (O3). We find that for individual BNSs at a network signal-to-noise ratio of 50, the systematic errors on \dl -- and hence \hubble -- are still smaller than the statistical uncertainties. 
The biases become more significant when we combine multiple events to obtain a joint posterior on \hubble. 
In the rather unrealistic case that the data around each detection is affected by the same CEs corresponding to the worst offender of O3, the true \hubble value would be excluded from the 90\% credible interval after we combine $\sim40$ sources. If instead, 10\% of the sources suffer from severe CEs, the true value of \hubble is included in the 90\% credible interval even after 100 sources.

\end{abstract}

\keywords{keywords}
\maketitle

\section{Introduction}
\label{Sec:Intro}
 The Hubble constant, \hubble, is the current expansion rate of the Universe and serves an important role in our understanding of the cosmic expansion history. However, there is currently a 4.4-$\sigma$ tension between the measurements of the early and the late Universe. Planck satellite’s observations of the cosmic microwave background anisotropies, assuming the standard flat cosmological model, lead to an inferred late-Universe measurement of $\hubble =67.36\pm0.54~\mathrm{km \ s}^{-1}\mathrm{Mpc}^{-1}$~\cite{DES:2017txv, Planck:2018vyg}. Direct measurements in the local universe lead to a different result: the SH0ES team measured $\hubble=74.03\pm1.42~\mathrm{km\ s}^{-1}\mathrm{Mpc}^{-1}$ using the Cepheid-supernova distance ladder~\cite{Riess:2016jrr, Riess:2018byc}, which is consistent with results from H0LiCOW using lensed quasars~\cite{Birrer:2018vtm,Wong:2019kwg}, and from the Hubble Space Telescope photometry and Gaia parallaxes~\cite{Riess:2020fzl}. The result from the Carnegie-Chicago Hubble program using the Tip of the Red Giant Branch method to calibrate the distances~\cite{Jang:2017dxn,Freedman:2019jwv,Kim:2020gai} is mostly consistent with the direct measurements, although it does not rule out the late-Universe measurements either.

Compact binary coalescences that emit both electromagnetical (EM) radiation and substantial gravitational waves (GWs) provide an independent method to measure \hubble and have the potential to resolve the above-mentioned tension~\cite{Schutz:1986gp,Holz:2005df,Dalal:2006qt,Nissanke:2009kt,Nissanke:2013fka,Chen:2017rfc,Hotokezaka:2018dfi,DES:2019ccw,Feeney:2020kxk}. Combining the luminosity distance \dl measurement from the GW observations and the Hubble flow velocity $v_H$ measurement from the EM observations leads to a direct estimation of \hubble, a method often referred to as the ``bright siren'' measurement. Since EM-bright GW sources detected by the second-generation GW detectors will be relatively local, we can use bright sirens to constrain only \hubble. Other cosmological parameters can be measured with methods that rely on the neutron star equation of state~\cite{Messenger:2011gi,DelPozzo:2015bna,Dietrich:2020efo}, dark sirens with galaxy catalogs~\cite{Schutz:1986gp,DelPozzo:2011vcw,LIGOScientific:2019zcs,LIGOScientific:2019zcs,DES:2019ccw,Gray:2019ksv,2019ApJ...871L..13F}, features in the mass distribution of binary neutron star mergers (BNSs) and binary black hole mergers~\cite{Taylor:2011fs,Farr:2019twy,Mastrogiovanni:2021wsd,Finke:2021aom}, or the spatial clustering scale of GW sources with known galaxies~\cite{Mukherjee:2020hyn,Diaz:2021pem}.

The observations of GW170817 by the Advanced LIGO and Virgo network (aLIGO-Virgo)~\cite{KAGRA:2013pob, LIGOsen,2015,PhysRevLett.123.231107}, the kilonova AT2017gfo and the gamma-ray burst GRB170817A served as the first demonstration of the bright-siren approach, leading to a measurement of $\hubble =70.0^{+12.0}_{-8.0}~\mathrm{km \ s}^{-1}\mathrm{Mpc}^{-1}$ ~\cite{LIGOScientific:2017adf, LIGOScientific:2017zic, LIGOScientific:2017vwq, LIGOScientific:2017ync, LIGOScientific:2018hze}. 
This method can be very powerful as we observe more such events. Reference~\cite{Nissanke:2013fka} projected a 5\% precision in the \hubble measurement after 15 BNSs with EM counterparts, and 1\% with 30 such sources. 

However, GW observations may suffer from underlying systematic biases in their estimate of \dl, and one needs to understand these biases before interpreting the resulting \hubble measurements. One such bias is the systematic error in the production and calibration of the detectors' primary data stream~\cite{Hall:2017off, Sun:2020wke,Sun:2021qcg}, referred to as calibration errors (CEs). Such errors are uncorrelated with the astrophysical event rate, evolve independently in each detector, and may be present in one or more detectors during several astrophysical events. CEs may bias the amplitude of the strain data in the same way over multiple observations, {and lead to a biased inference of \hubble.}

Previous studies have estimated the impact of typical values of CEs on individual events observed during the aLIGO-Virgo's first and second observing runs ~\cite{Vitale:2011wu,Payne:2020myg, Vitale:2020gvb}, attempted to refine the estimates of CEs using detected~\cite{Essick:2019dow} or expected astrophysical signals~\cite{Schutz:2020hyz}, or derived analytically how CEs can impact the leading-order tidal parameter in next-generation GW detectors~\cite{PhysRevD.105.082002}. Our study instead uses large CEs experienced during atypical times in the aLIGO-Virgo's third observing run (O3) to probe their worst-case impact on astrophysical parameter estimation (PE) for both the single-event characterization and the joint inference of \hubble. %

More specifically, we simulate a collection of BNS detections and introduce into the GW data stream of each signal artificial CEs that follow six cases of particularly large CEs from the Advanced LIGO detectors at Hanford (LHO) or Livingston (LLO) during O3~\cite{Sun:2020wke, Sun:2021qcg}. We pick the CE realization that leads to the worst bias in the astrophysical measurements of \dl, apply it to varying fractions of 100 simulated BNSs, and infer \hubble to explore the progressive impact of large CEs. 
In Sec.~\ref{Sec:Method}, we will discuss the method we use to produce the miscalibrated data stream and to perform PE. In Sec.~\ref{Sec:Results}, we report how CEs impact the \dl results for individual events, as well as the \hubble results when we combine multiple sources.

\section{Method}
\label{Sec:Method}
We summarize how we simulate the data stream in Sec.~\ref{Subsec:Sim}, how we define CEs and artificially add them to the data stream in Sec.~\ref{Subsec:SysError}, and how we perform PE and infer \hubble in Sec.~\ref{Subsec:astroPE}.

\subsection{Simulations}
\label{Subsec:Sim} 

We simulate 5000 BNSs with uniform-in-cosine inclinations and uniformly distributed sky locations. Each event is assigned a \dl randomly drawn from a uniform-in-comoving-volume distribution, with the maximum \dl set at 600 Mpc, larger than the horizon~\footnote{The horizon is estimated for events with a network SNR of 12.} of the aLIGO-Virgo network at the design sensitivity~\cite{KAGRA:2013pob,LIGOsen}. We randomly draw 100 events with network matched-filter signal-to-noise ratios (SNRs) above 12 to form our set of EM-bright GW events.

We simulate nonspinning BNSs with component masses $m_1=2M_\odot$ and $m_2=1.5M_\odot$ with the phenomenological waveform model \textsc{IMRPhenomPv2}~\cite{Husa:2015iqa, Khan:2015jqa, Hannam:2013oca, Smith:2016qas}, which does not model neutron star matter effects. Our choice of $m_1$ and $m_2$ are potential masses for neutron stars. Since the BNSs contain a significant number of cycles that allow us to constrain the masses very well, the specific choice within the allowed range will not impact our study.

We use the same waveform model during PE in order to avoid any systematics caused by waveform mismatch.\footnote{Previous papers have looked into waveform systematics for BNSs, for example, Refs.~\cite{Ohme:2013nsa,Berry:2014jja,Samajdar:2018dcx}.} We assume an EM counterpart yielded an exact redshift measurement for each event. Since the uncertainties on the sky localization from the EM measurements will be much smaller than the typical uncertainties on the GW measurement, we assume the sky positions of the sources are exactly known from the EM observations prior to the $H_0$ analysis. We also disregard effects from the peculiar velocity of the sources, as most of the events will be farther than 150 Mpc. For a peculiar velocity of 200 km/s, the resulting biases in the $H_0$ measurements are less than 2\%~\cite{2015MNRAS.450..317C,2018ApJ...859..101S}. %

We produce Gaussian noise ${\bf n}$ colored by the aLIGO-Virgo design sensitivities~\cite{LIGOsen}. We then project the signal, which is the sum of the modeled waveforms and the noise, to each detector of the network to obtain the data stream.

\subsection{Systematic calibration errors}
\label{Subsec:SysError}

Each of the current aLIGO-Virgo detectors is a dual-recycled Fabry-P\'erot Michelson laser interferometer (IFO)~\cite{buikema2020sensitivity,acernese2014advanced}. Its data stream \ifod is obtained from a voltage signal, \ed, measured from the output laser power incident on a photodetector. The process of converting \ed into \ifod is referred to as calibration~\cite{Viets_2018}. For each detector at any given time, the conversion from \ed to \ifod is made by a complex-valued, frequency-dependent response function, $R(f;t)$~\footnote{\ifod and \ed are implicit functions of $f$ and $t$.}, 
\begin{equation}
    \ifod = R_\mathrm{IFO}(f;t)~\ed,
\end{equation} 
where $f$ is frequency and $t$ is time. The model for the response function, $\rmodIfo(f;t)$, is constructed from the expected behavior of the detectors, coupled with supporting measurements of the model parameters as described in Refs.~\cite{PhysRevD.95.062003,Tuyenbayev:2016xey}.
Imperfections in $\rmodIfo(f;t)$, and thus in \ifod, are referred to as calibration errors, or CEs~\cite{Hall:2017off, Sun:2020wke, Sun:2021qcg}.
The CEs, $\eta$, may be represented by the ratio of the true response function, $\rtrueIfo(f;t)$, and the model, $\rmodIfo(f;t)$, by \footnote{$\eta_{\rm IFO}(f;t)$ is referred to as $\eta_R(f;t)$ in Refs.~\cite{Sun:2020wke, Sun:2021qcg}.}%
\begin{equation}
    \eta_{\rm IFO}(f;t)= \rtrueIfo(f;t)/\rmodIfo(f;t).
\end{equation} 
In the ideal case, $\rmodIfo(f;t)$ will be identical to $\rtrueIfo(f;t)$, and $\eta_{\rm IFO}(f;t)$ becomes a frequency-independent constant with unity magnitude and zero phase. In reality, $\eta_{\rm IFO}(f;t)$ is usually a function of frequency and time. 
Note that the transferring laser power calibration uncertainty from the US
National Institute of Standards and Technology (NIST), which impacts the fiducial displacements of the test masses, appears in all detectors as a subdominant time-independent, frequency-independent common offset~\footnote{The common offset on the fiducial displacements is at the level of $<$ 0.5\% throughout O3~\cite{Sun:2020wke,Sun:2021qcg,Bhattacharjee_2021} and has negligible impact on distance measurement of the events compared to CEs caused by imperfections in the calibration models~\cite{Vitale:2020gvb}.} and has been accounted for during PE. Since $\rmodIfo(f;t)$, $\rtrueIfo(f;t)$ and $\eta_{\rm IFO}(f;t)$ always depend on $f$ and $t$ in our case and evolve independently in every detector, we will drop the $(f;t)$ arguments and the IFO subscript henceforth unless we need to specify $f$, $t$, or IFO. 

While \rmod is known, \rtrue may only be inferred by direct measurements that are invasive for observations and would reduce the duty cycle of the detector. Instead, at any given reference time $t_k$, the parameters in \rtrue are numerically sampled to create $10^4$ realizations representing the estimated probability distribution of $\{\eta(t_k)\}$, with relatively large uncertainty. The value of $\eta$ may vary over time, either discretely with changes made in the detector's control system between different configuration periods (typically months apart), or with drifts in the detector's alignment or thermal state. In O3 when the detector was stable in the observing mode, the probability distribution of $\eta$ for each detector was estimated at discrete times with hourly cadence. The slow time-dependent variation is appropriately tracked and accounted for in the hourly estimated $\{\eta(t_k)\}$ in most cases. Over a particular observing period of static configurations, we can choose an arbitrary reference time to represent the typical CE distribution of this period, referred to as $\{\eta^\mathrm{typ}\}_k$.

The entire detector's response is checked weekly and monitored continuously at a few select frequencies, to ensure coverage of any unexpected situations. When issues are found in the detector behavior, additional measurements may also be made to improve the estimation for $\{\eta(t_{k})\}$ retroactively. If at a time $t_i$, the updated $\{\eta(t_i)\}$ is significantly different, e.g., by $\sim 10\%$ in magnitude and/or $\sim 10$ degrees in phase at any given frequency, from the typical behavior  estimated at a nearby time, $\etaTyp_i$, we refer to these outliers as $\{\eta^\mathrm{out}(t_i)\}$ (henceforth $\etaOut_i$). During O3, six such outliers have been identified in either LLO or LHO. For example, in Fig.~\ref{Fig.distri_1263066740_BNS}, we compare the median and 1-$\sigma$ boundaries of the magnitude and phase of the $\etaOut_{6,\mathrm{LLO}}$ distribution against its corresponding $\etaTyp_{6,\mathrm{LLO}}$ distribution. We include plots for the distributions of $\etaOut_i$ and $\etaTyp_i$ for $i=1,...,5$ in Appendix~\ref{App.rdis}. 

In reality, if there happens to be an astrophysical signal at a time with outlier CEs, an updated estimate of the true CE distribution may not be readily available at the time of PE analyses, or the CEs are not updated in the rare case that an outlier is not detected. Thus, $\etaTyp$ of this time will be used during PE. This is the scenario we investigate in this study: when the actual CEs are very different from the CEs known and used at the time of PE.
\begin{figure}
\centering
\begin{subfigure}[b]{0.48\textwidth}
   \includegraphics[width=1\linewidth]{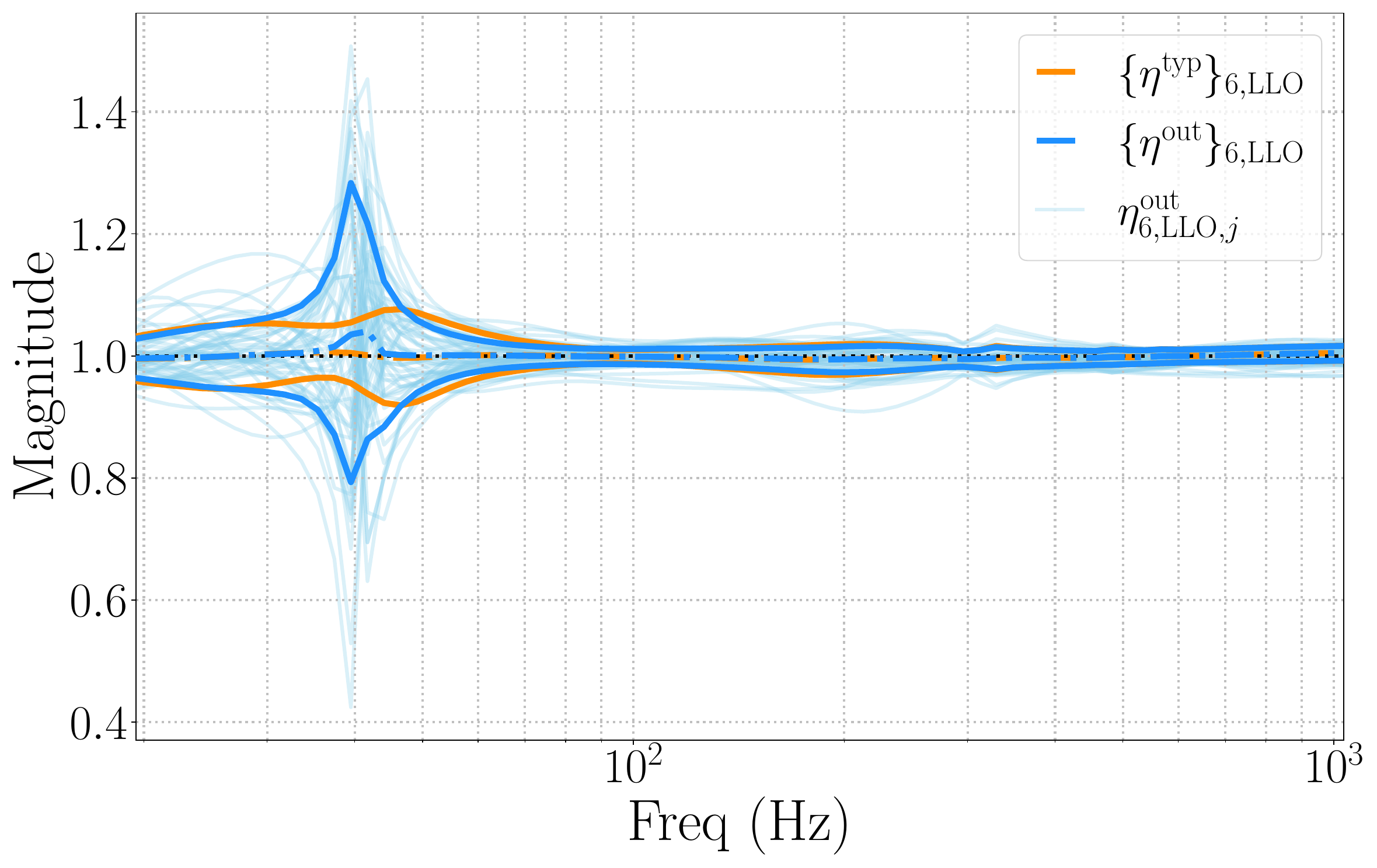}
   \vspace{-1.5\baselineskip}
   \label{Fig.distributions_mis_amps_1263066740_L1} 
\end{subfigure}
\begin{subfigure}[b]{0.48\textwidth}
   \includegraphics[width=1\linewidth]{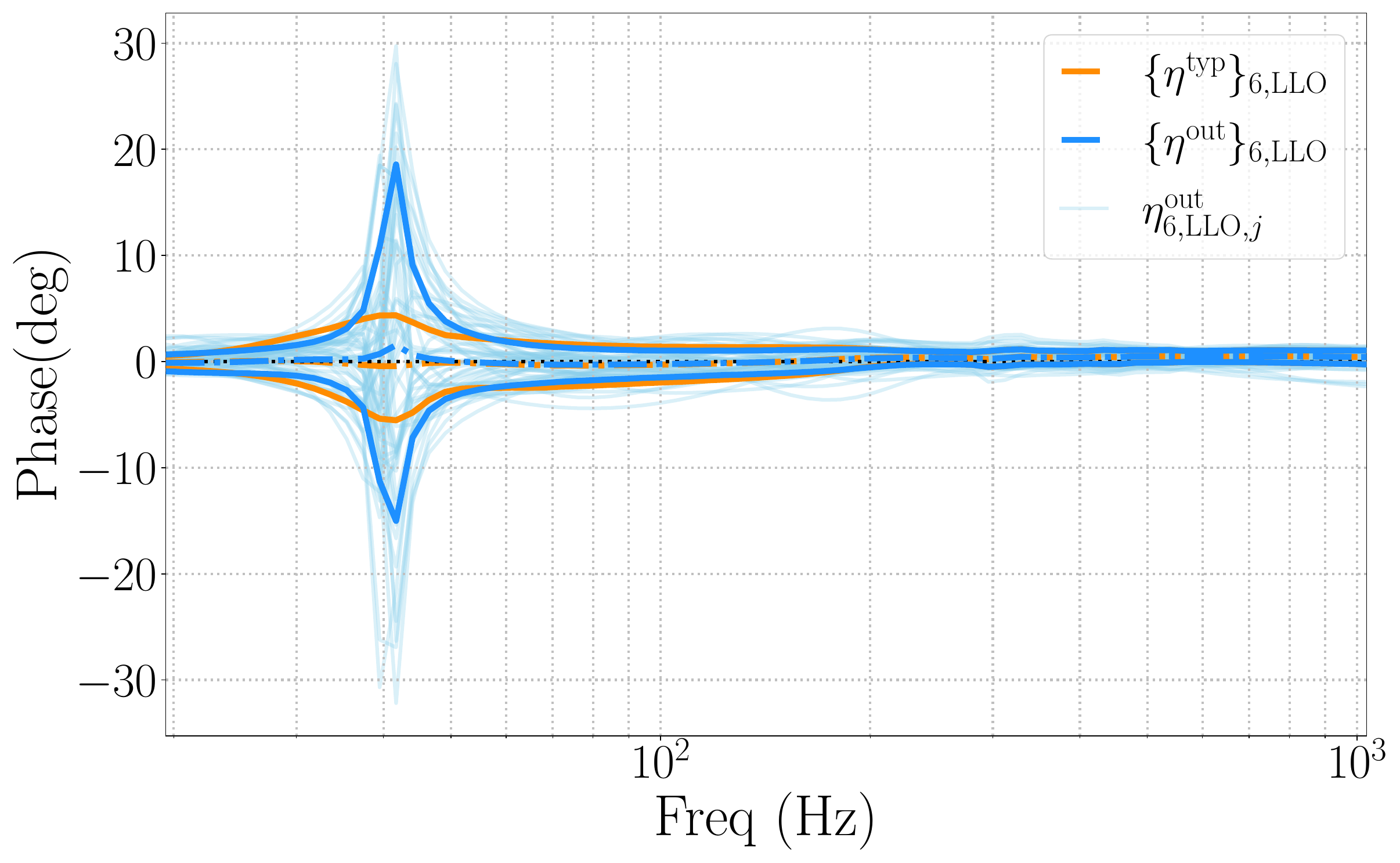}
   \vspace{-1.5\baselineskip}
   \label{Fig.distributions_mis_pha_1263066740_L1}  
\end{subfigure}
\caption{The median (dash-dotted) and the 1-$\sigma$ bounds (solid, thick) of one example of a large CE, $\etaOut_{6,\mathrm{LLO}}$ from O3, in magnitude (top) and phase (bottom) as a function of frequency (dark blue). The thin, light blue curves represent some of the individual numerically estimated realizations, $\eta^\mathrm{out}_{6,\mathrm{LLO},j}$ during this time. The orange curves are the corresponding typical distribution $\etaTyp_{6,\mathrm{LLO}}$.}
\label{Fig.distri_1263066740_BNS}
\end{figure}
To study this scenario, we apply frequency-dependent, artificial outlier CEs, $\miscal_i$, to the data streams $\boldd_\mathrm{LLO}$ or $\boldd_\mathrm{LHO}$.  For each of the six cases of study, the outlier CEs $\miscal_i$ are applied to only \emph{one} of the aLIGO detectors. The selection process for $\miscal_i$ is as follows. We desire $\miscal_i$ to have the most impact on the amplitude of the detector data, to thus have the most influence on the estimated $\dl$. We therefore pick the worst realization from each $\etaOut_i$ distribution, where ``worst'' is defined by the impact, $\mathcal{D}_{i,j}$, of applying this error:
\begin{equation}
\label{Eq.IntDiff}
\mathcal{D}_{i,j} \equiv \int^{f_\mathrm{high}}_{f_\mathrm{low}} \frac{|\etaOut_{i,j}| - \widetilde{|\etaTyp_i|}}{\sqrt{S_n(f)}}\mathrm{d}f.    
\end{equation}
In this definition, $j$ indexes an individual realization from each $\etaOut_i$ distribution. At each frequency, we take the difference between the magnitude of the $j^{\rm th}$ realization of $\etaOut_{i,j}$ (a thin blue curve in Fig.~\ref{Fig.distri_1263066740_BNS}), and the median of the magnitudes of samples in $\etaTyp_i$, $\widetilde{|\etaTyp_i|}$ (the dot-dashed orange curve in Fig.~\ref{Fig.distri_1263066740_BNS}). The detector's amplitude spectral density is denoted by $\sqrt{S_n(f)}$, for which we use aLIGO's design sensitivity~\cite{KAGRA:2013pob,LIGOsen}. %
The frequency limits of the integral, $f_\mathrm{high}$ and $f_\mathrm{low}$, have been chosen as 20 and 1024Hz, respectively, with a frequency resolution of 0.25Hz. Given that each magnitude, $|\etaOut_{i,j}|$ or $\widetilde{|\etaTyp_i|}$, is frequency-dependent and can be smaller or larger than unity, the impact, $\mathcal{D}_{i,j}$, may be either positive or negative at any given frequency.
We select $\miscal_i$ to be the $\etaOut_{i,j}$ that maximizes $|\mathcal{D}_{i,j}|$,\footnote{We performed the full PE analysis with the curve that maximizes $\mathcal{D}_{i,j}$ and $|\mathcal{D}_{i,j}|$, and observed bigger biases in the PE results for maximized $|\mathcal{D}_{i,j}|$. }
\begin{equation}
\label{Eq.pickMiscal}
\miscal_i \equiv \underset{j}{\max} |\mathcal{D}_{i,j}|.
\end{equation}
We apply $\miscal_{i}$ to miscalibrate the data, i.e., the sum of the noise and the modeled waveform, as,
\begin{equation}
\label{Eq.miscal}
\dmis_i = \miscal_i~\boldd.
\end{equation}
The noise, as part of \boldd, will thus also be scaled by $\miscal_i$. The resulting amplitude spectral density of each detector is $\miscal_i\sqrt{S_n(f)}$. In the case of $i=6$, for example, we apply $\miscal_{6,\mathrm{LLO}}$ to miscalibrate the LLO data.

For the other aLIGO detector, in this case LHO, we select a curve $\eta_{6,\mathrm{LHO}}$ that lies within the 1-$\sigma$ credible interval (CI) of $\etaTyp_{6,\mathrm{LHO}}$ to miscalibrate the data, in the same way as Eq.~\eqref{Eq.miscal}. No CEs are added to the aVirgo data, since the full $\eta$ distributions of aVirgo are not available at the time of writing.
$\{\miscal_{6,\mathrm{LLO}},\eta_{6,\mathrm{LHO}},1\}$ forms the set of curves to miscalibrate the data for scenario number 6. Figure~\ref{Fig.mis_1263066740_BNS} shows the comparison between the selected $\miscal_{6,\mathrm{LLO}}$ (green) and the median and 1-$\sigma$ bounds of the $\etaTyp_{6,\mathrm{LLO}}$ distribution (orange) from Fig.~\ref{Fig.distri_1263066740_BNS} in magnitude and phase. 

We also prepare a separate set of \emph{control runs} in which we do not add \emph{any} CE in any of the detectors. When comparing PE results of the miscalibrated runs and the control runs, we can observe biases caused exclusively by the large added CEs in the former. 

\begin{figure}
\centering
\begin{subfigure}[b]{0.48\textwidth}
   \includegraphics[width=1\linewidth]{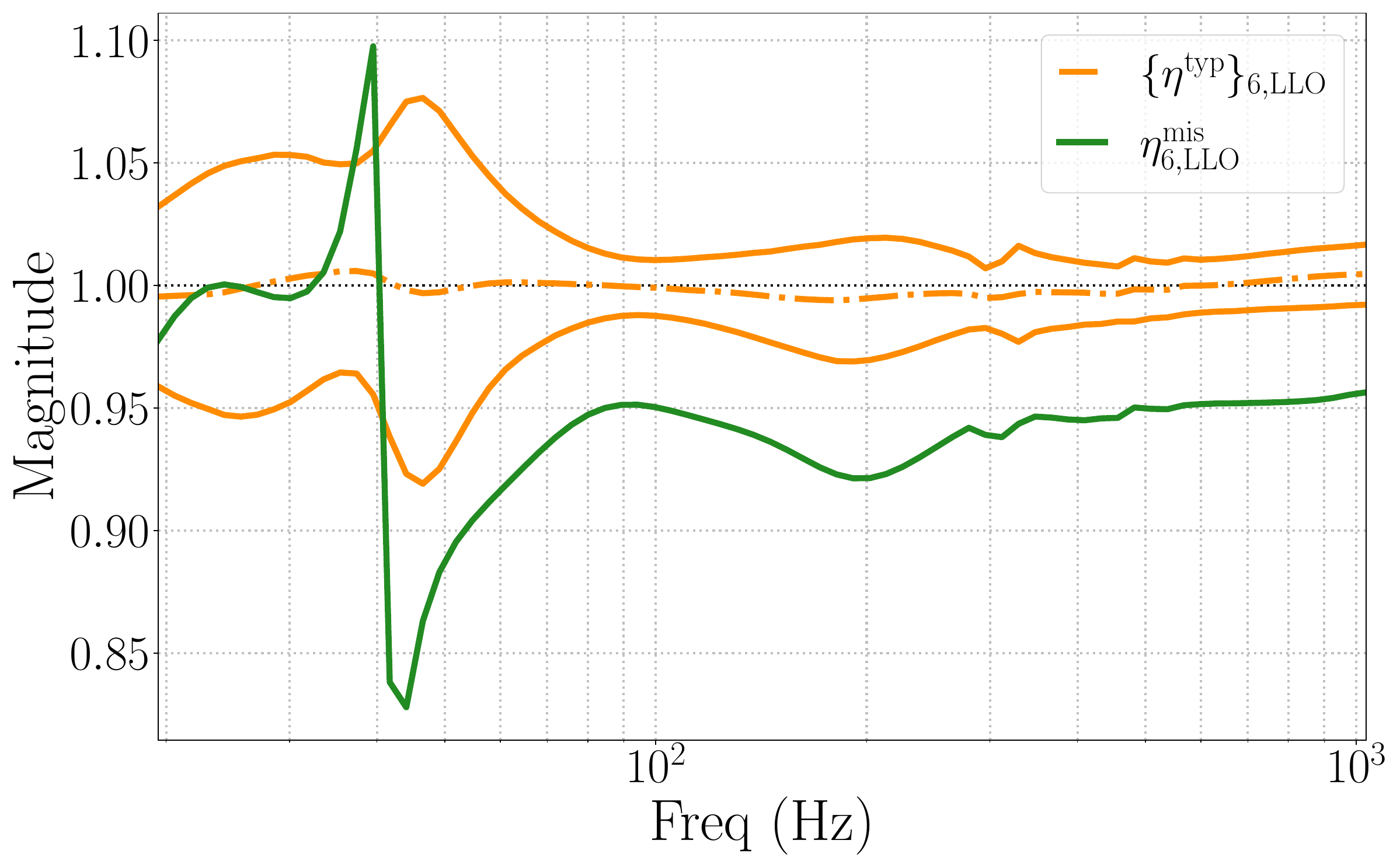}
   \vspace{-1.5\baselineskip}
   \label{Fig.mis_amp_1263066740_L1_BNS} 
\end{subfigure}
\begin{subfigure}[b]{0.48\textwidth}
   \includegraphics[width=1\linewidth]{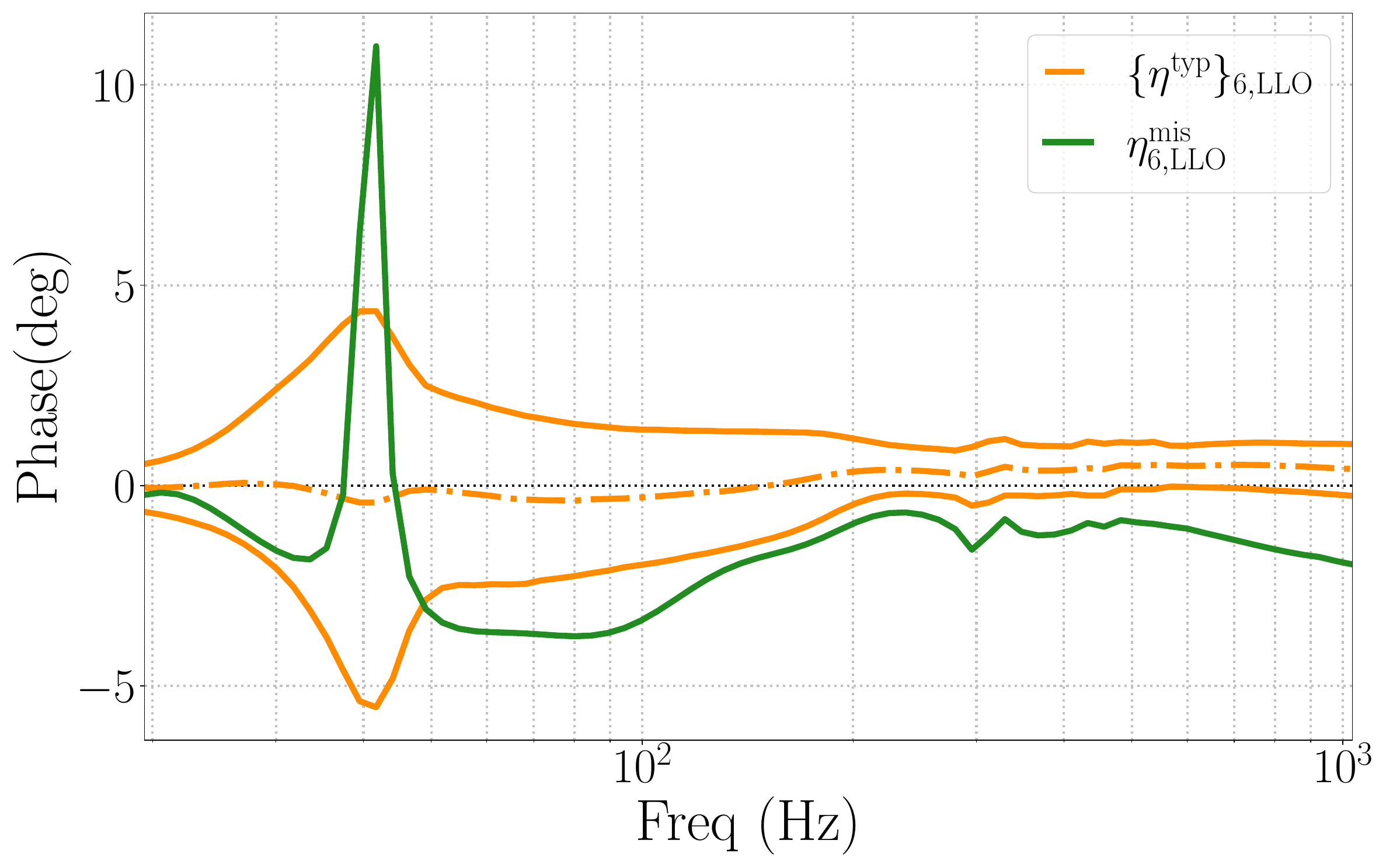}
   \vspace{-1.5\baselineskip}
   \label{Fig.mis_phase_1263066740_L1_BNS}  
\end{subfigure}
\caption{$\miscal_{6,\mathrm{LLO}}$ (green) shown in magnitude and phase (deg) as a function of frequency (Hz). Also plotted are the 1-$\sigma$ bounds and the median of typical CE distributions, $\etaTyp_{6,\mathrm{LLO}}$ (orange). }
\label{Fig.mis_1263066740_BNS}
\end{figure}

Next, we select one CE realization that leads to the most significant bias in the \dl likelihood, and apply it to 100 segments of data, each containing a different simulated BNS event, as described in Sec.~\ref{Subsec:Sim}. To select the desired CE realization, we first consider a single BNS event with a network SNR of 50, an inclination of 30$^\circ$, and a sky location right above LLO in the sky. The choice of SNR 50 is made to strike a compromise between an SNR high enough that systematics are important, but low enough that it might occur in the next observing runs.

We add each of the six distinctive sets of CEs to a data chunk containing this BNS, and select the realization that leads to the most significant bias in the \dl likelihoods. As we only consider single-event results in this part of the study, we need to eliminate random effects of noise realizations. We work with the zero-noise realization~\cite{Vallisneri:2007ev} where the noise ${\bf n}$ has mean of ${\bf 0}$ and standard deviation equal to the detector's power spectral density. Lastly, we infer \hubble from the 100 events. 

\subsection{Parameter estimation and inference of \hubble}
\label{Subsec:astroPE} 

We perform Bayesian inference~\cite{Veitch:2014wba,LIGOScientific:2016vlm} to obtain the likelihood $\llhd({\boldsymbol {d}}|\boldth,\hubble,\mathcal{H})$, where $\boldth$ is the set of astrophysical parameters, under model $\mathcal{H}$. We write \hubble explicitly because it is the hyperparameter that we are interested in. Since not all events have equal chances of detection, we follow Refs. ~\cite{Chen:2017rfc,Vitale:2020aaz} to account for selection effects, and write the detectable sources as a function of $\hubble$, denoted by $\beta(\hubble)$. We can obtain the likelihood $\llhd(\boldd|\hubble,\mathcal{H})$ by integrating over all astrophysical parameters and account for the selection effects,
\begin{equation}
\label{Eq.H0Likelihood}
\llhd(\boldd|\hubble,\mathcal{H})= \frac{\int \mathrm{d}\boldth \llhd(\boldd | \boldth,\hubble,\mathcal{H})\pi(\boldth|\hubble,\mathcal{H})}{\beta(\hubble)} ,
\end{equation}
where $\pi(\boldth|\hubble,\mathcal{H})$ is the prior.  Since all sources in our simulations have the same source-frame masses, the selection function only needs to be calculated once.

We can obtain the posterior for \hubble by applying a uniform prior of $\pi(\hubble | \mathcal{H})= [20,150] \mathrm{km\ s}^{-1}\mathrm{Mpc}^{-1}$, 
\begin{equation}
\label{Eq.BayesTheorem}
p(\hubble|{\boldsymbol {d}},\mathcal{H})  = {\pi(\hubble|\mathcal{H})\llhd({\boldsymbol {d}}|\hubble,\mathcal{H})}
\end{equation}
As all of the 100 BNSs are independent,  their joint likelihood can be calculated by simply multiplying the likelihoods of each event~\cite{Zimmerman:2019wzo}, 
\begin{equation}
\label{Eq.LikelihoodSum}
\begin{aligned}
\llhd({\boldsymbol {d}}|\hubble,\mathcal{H})= \underset{i=1}{\overset{100}{\Pi}} \llhd_i({\boldsymbol {d}}|\hubble,\mathcal{H}).
\end{aligned}
\end{equation}

One caveat is that our PE analysis adopts a standard, uniform-in-individual-masses prior with additional bounds on the chirp mass $\mchirp\equiv (m_1m_2)^{3/5}/(m_1+m_2)^{1/5}$ and mass ratio, while our simulated events all have the same masses, and the function of event detectability in Eq.~\eqref{Eq.H0Likelihood} is calculated assuming so. This assumption does not fully capture the intrinsic relationship between source-frame masses, redshift, and cosmological parameters like $H_0$. As a result, it may introduce a bias, as a uniform mass prior might give equal weight to events that are actually less likely given a particular cosmology. Furthermore, the spectral siren method~\cite{PhysRevLett.129.061102} relies on this intrinsic relationship and utilizes the mass spectrum when calculating cosmological constraints. Nevertheless, in this analysis, the same effect caused by the assumption will be present in the results for both the miscalibrated runs and the control runs, thus not affecting the differential behaviors between runs. Recently, model agnostic methods have also been developed 
to mitigate the bias introduced by fixed mass spectrum ~\cite{farah2024needknowastrophysicsfreegravitationalwave,hernandez2024gapsbumpsspectralsiren}.

In all PE analyses to date, we account for known CEs by marginalizing over our best estimates of the CE distribution through two approaches: the Spline interpolation method~\cite{Spline} and the physiCal method~\cite{Payne:2020myg,Vitale:2020gvb}. Both methods treat the CEs as independent in each detector. The Spline method models $\eta$ by fitting a cubic spline polynomial at a small number of logarithmically spaced frequencies $\{ f_m\}$. At each frequency, the prior on the magnitude and phase is a Gaussian distribution with the same mean and standard deviation as those of $\{\eta^\mathrm{typ}(f_m)\}$. The recently developed physiCal method~\cite{Payne:2020myg, Vitale:2020gvb} is more computationally efficient and physically motivated, and estimates the physical parameters in the models for $\eta$ along with $\boldsymbol{\theta}$ during PE. PhysiCal directly draws samples from \etaTyp to form the prior.

In this study, we are interested in scenarios where large CEs are {\it not} fully captured and thus not accounted for in PE. We assume that we do not know the true error distribution $\etaOut_i$, but only know and use $\etaTyp_i$ as the prior for physiCal. %
Similarly, we utilize the medians and bounds of 1-$\sigma$ CIs of $\etaTyp_i$ as priors for the Spline method. As mentioned in Sec.~\ref{Subsec:SysError}, since the full $\{\eta\}$ distributions of aVirgo were not available, we only adopt the Spline method to marginalize CEs in aVirgo.

\section{Result}
\label{Sec:Results}

First, we present \dl~posteriors with a flat prior to illustrate the likelihood distributions (referred to as ``likelihood'' henceforth for simplicity) when we apply the six sets of $\miscal$ (selected using the method described in Sec.~\ref{Subsec:SysError}) to a single BNS event with a network SNR of 50. The likelihoods for the miscalibrated runs are plotted as green kernel density plots in Fig.~\ref{Fig.dist_BNS_SNR50_Split_ll}, and the results for the corresponding control run are plotted in orange. We report the normalized difference $\Delta \dl = (D_{L,\mathrm{med}}-D_{L,\mathrm{true}})/D_{L,\mathrm{true}}$, between the true value $D_{L,\mathrm{true}}$ and the median of the recovered likelihoods, $D_{L,\mathrm{med}}$, in Table~\ref{Table.dl bias}. 

\begin{figure}
\centering
\includegraphics[width=\linewidth]{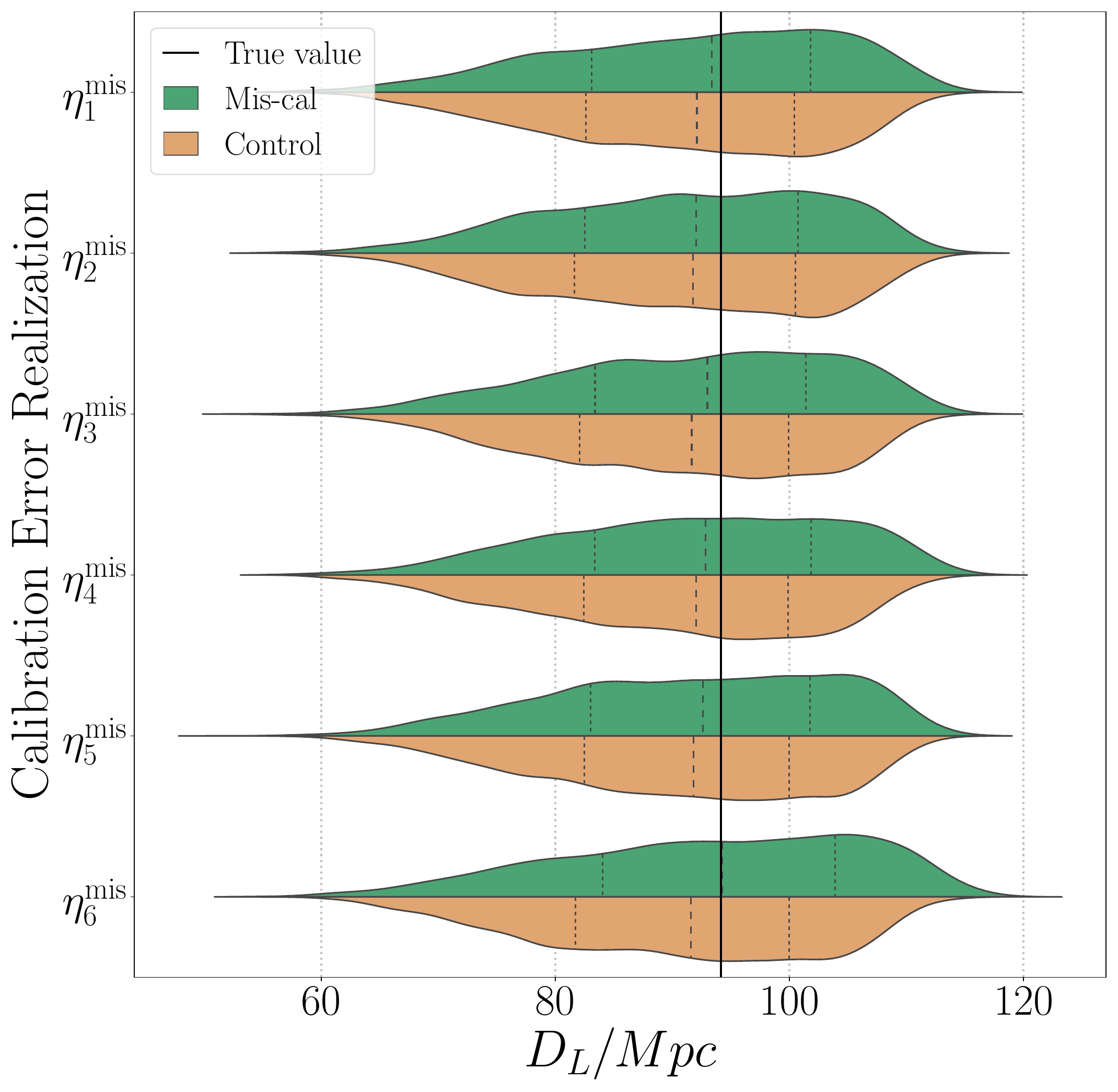}
\caption{\dl likelihood for the six scenarios, miscalibrated (green) vs. control (orange) runs, the vertical dashed lines mark the 25\%, 50\%, and 75\% percentiles.}
\label{Fig.dist_BNS_SNR50_Split_ll}
\end{figure}

\begin{table}[t]
\centering
\begin{tabularx}{0.90\linewidth}{b|bb}
\toprule\toprule
CE realization & miscalibrated & Control \\
\midrule
$\miscal_1$&$-1.0\%$&$-2.5\%$\\
\midrule
$\miscal_2$&$-2.3\%$&$-2.3\%$\\
\midrule
$\miscal_3$&$-1.2\%$&$-2.1\%$\\
\midrule
$\miscal_4$&$-0.9\%$&$-2.3\%$\\
\midrule
$\miscal_5$&$-1.5\%$&$-2.5\%$\\
\midrule
$\miscal_6$&$0.5\%$&$-2.6\%$\\
\bottomrule\bottomrule
\end{tabularx}
\caption{$\Delta \dl$ in the likelihoods for the physiCal runs with and without large CEs. }
\label{Table.dl bias}
\end{table} 

The results from using the physiCal and Spline methods agree very well. Thus, we only show the results from the physiCal methods in Fig.~\ref{Fig.dist_BNS_SNR50_Split_ll} and Table~\ref{Table.dl bias}, and the ones from the Spline method in Appendix~\ref{App.Spline}. Compared with the control runs, where the offsets are between $-2.1\%$ to $-2.6\%$, $\miscal_6$ leads to the most significant differences in the \dl likelihoods. We note the control runs all show a negative $\Delta\dl$ due to the well-known correlation between \dl and inclination~\cite{Chen:2018omi}. Since all sources in Fig.~\ref{Fig.dist_BNS_SNR50_Split_ll} have an inclination $\iota$ of 30$^\circ$, but the inclination prior follows $\sin\iota$, we expect an offset towards larger inclination values, where the prior is larger, and thus smaller distances. 

Next, we apply $\miscal_6$ to the data of 100 BNSs. This offset due to the inclination degeneracy will \emph{no longer} be present in the analysis since the inclinations are drawn from a uniform-in-cosine distribution (effectively $\sin\iota d\iota$), the same as the prior.

In Fig.~\ref{Fig.h0_partlyMiscaled}, we vary the fraction of miscalibrated BNSs and show the posterior distribution of the dimensionless $h_0$, defined as $\hubble = h_0 100~\mathrm{km \ s}^{-1}\mathrm{Mpc}^{-1}$. We apply $\miscal_6$ to $x\%$ of the BNSs, while the other events do not suffer from any miscalibration. The joint $h_0$ posterior shifts towards smaller values as the miscalibrated fraction increases. The posterior excludes the true value of ${h_0= 0.679}$ from the 90\% CI when the data of more than 50\% BNSs are miscalibrated. 

\begin{figure}
\centering
\includegraphics[width=\linewidth]{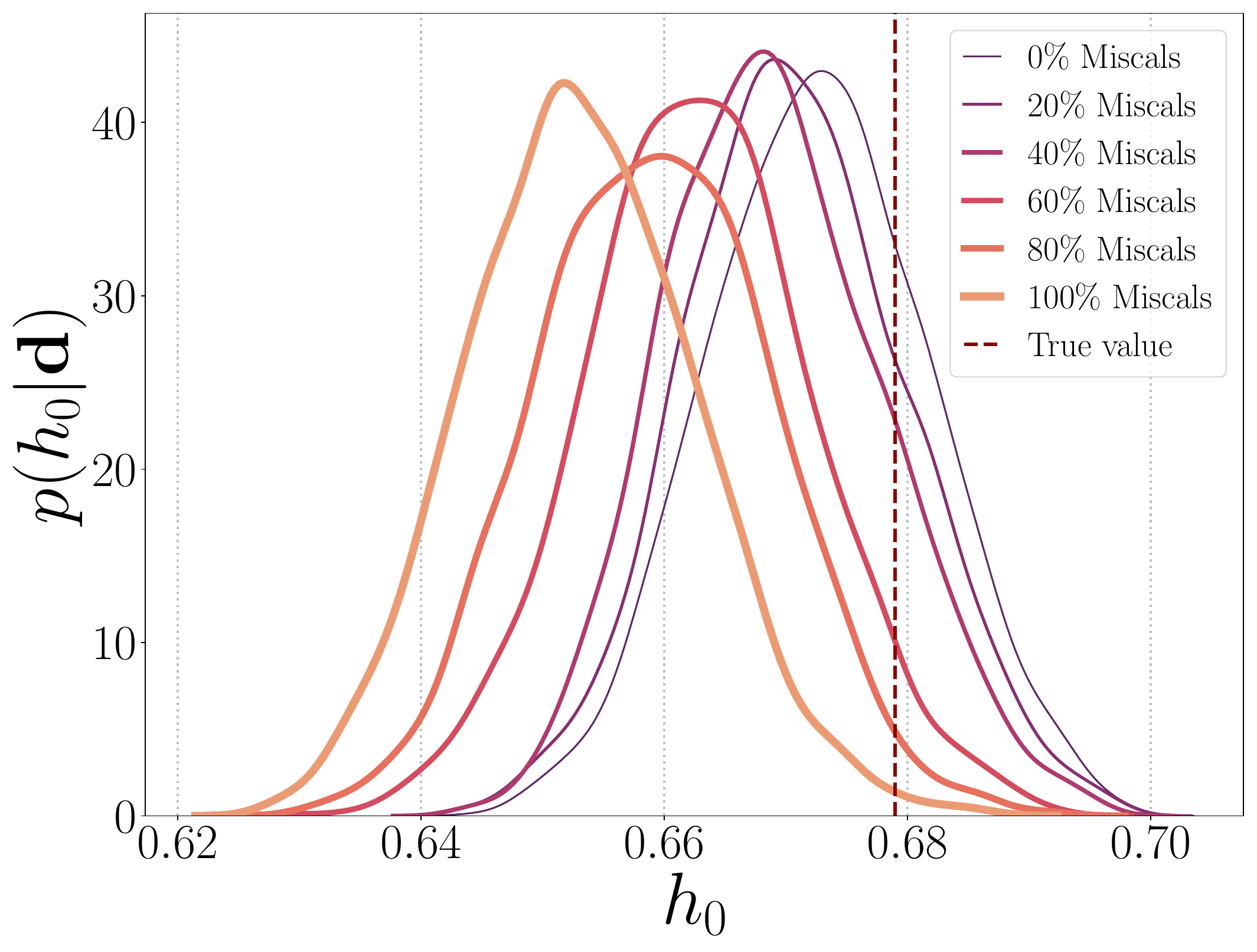}
\caption{$h_0$ posteriors for different fractions of miscalibration. Thicker curves indicate results with more BNS events miscalibrated.}
\label{Fig.h0_partlyMiscaled}
\end{figure}

Additionally, we vary the total number of detected events. 
We randomly draw $n$ detections without repetition for $\lfloor 100/n \rfloor$ trials (i.e., rounding down $100/n$ to an integer), and randomly miscalibrate $x\%$ of the events in each trial. We calculate the median and the bounds of the 90\% CI for each trial. In Fig.~\ref{Fig.h0_nEvents_cusRange}, for each pair of $\{n,x\}$, the green points are medians of the $\lfloor 100/n \rfloor$ medians from each trial, and the error bars are the medians of the 90\% CI bounds from each trial. 
The orange points and error bars indicate the results of the control runs.\footnote{The medians of the control runs for under 30 events jump above and below the true value. As we increase $n$, there are fewer trials, and our results show more dependence on specific noise realizations. In this specific case of noise realization, the median falls below the true value. If we apply different noise realizations, the results will jump around the true value like the trials with fewer events.}

With 100\% of the detected events miscalibrated, the joint $h_0$ posterior excludes the true value from its 90\% CI after 50 detections or more. 
With 50\% miscalibrated, the posterior only starts to exclude the true $h_0$ after more than 90 events. When 10\% miscalibrated, the posterior includes the true $h_0$ even after 100 detections. 

\begin{figure}
\centering
\begin{subfigure}[b]{0.48\textwidth}
   \includegraphics[width=\linewidth]{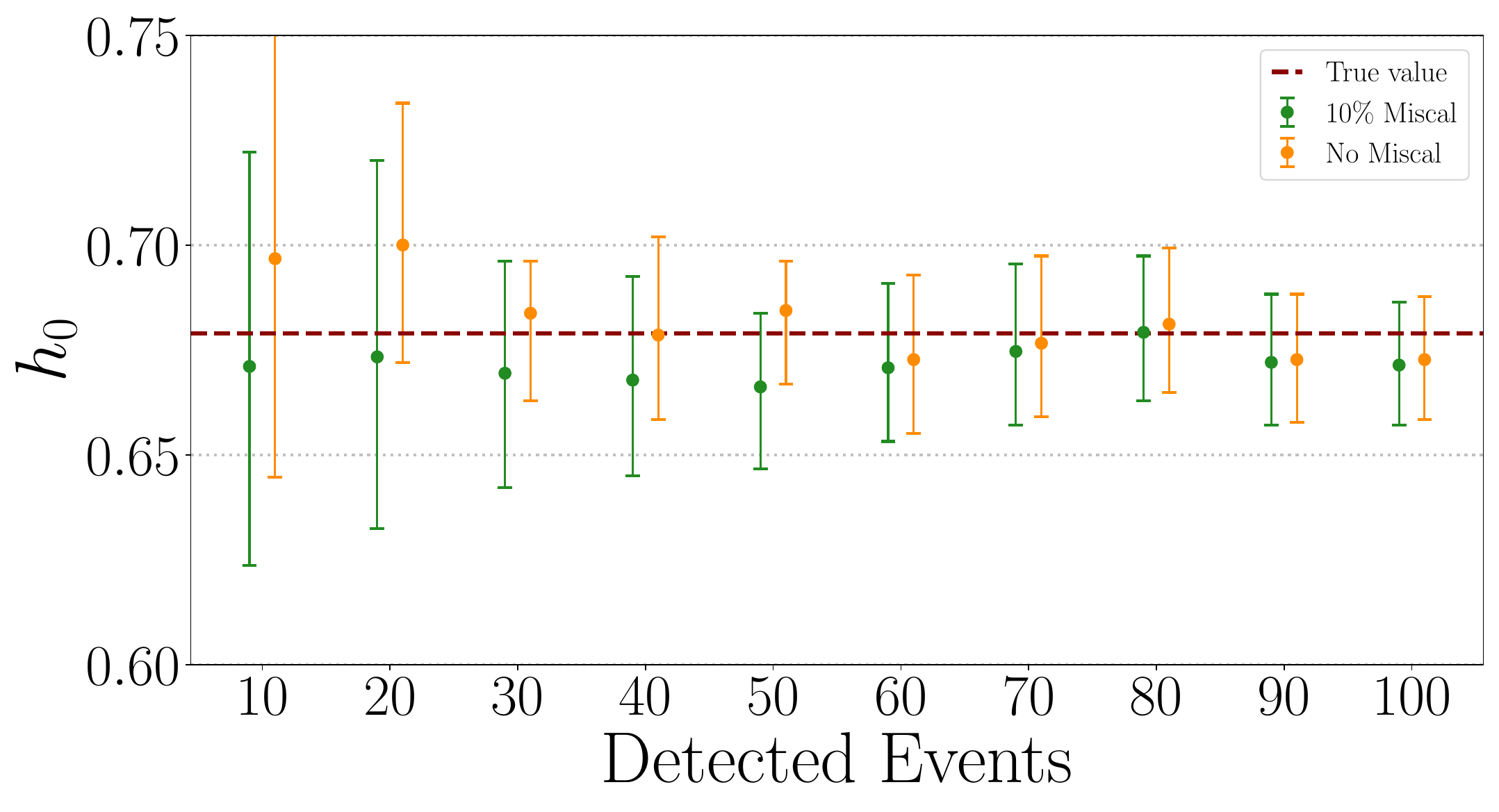}
   \vspace{-1.5\baselineskip}
   \caption{10\% of the events miscalibrated.}
   \label{Fig.h0_nEvents_cusRangemiscal0.1}  
\end{subfigure}
\begin{subfigure}[b]{0.48\textwidth}
   \includegraphics[width=\linewidth]{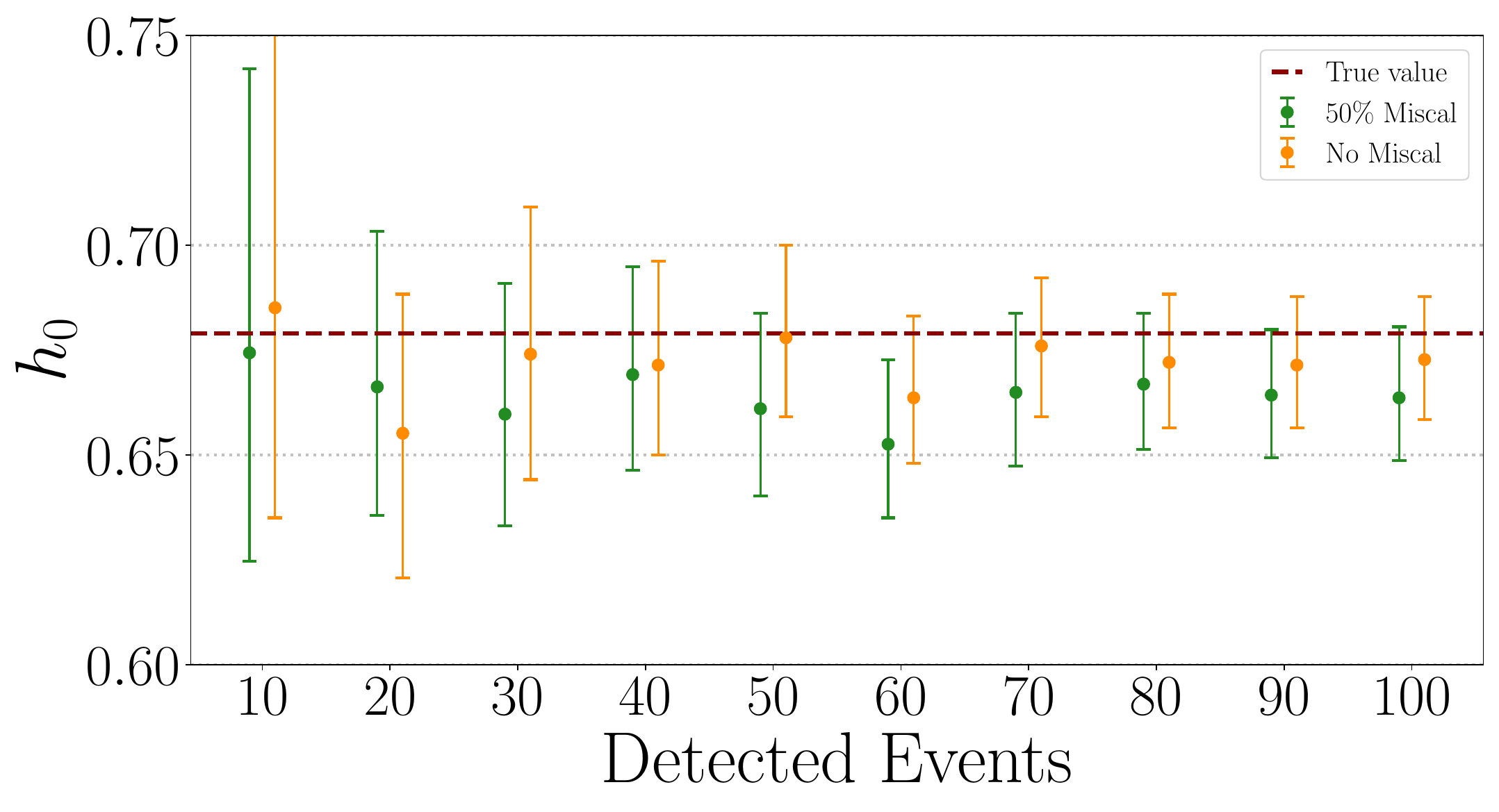}
   \vspace{-1.5\baselineskip}
   \caption{50\% of the events miscalibrated.}
   \label{Fig.h0_nEvents_cusRange_miscal0.5}  
\end{subfigure}
\begin{subfigure}[b]{0.48\textwidth}
   \includegraphics[width=\linewidth]{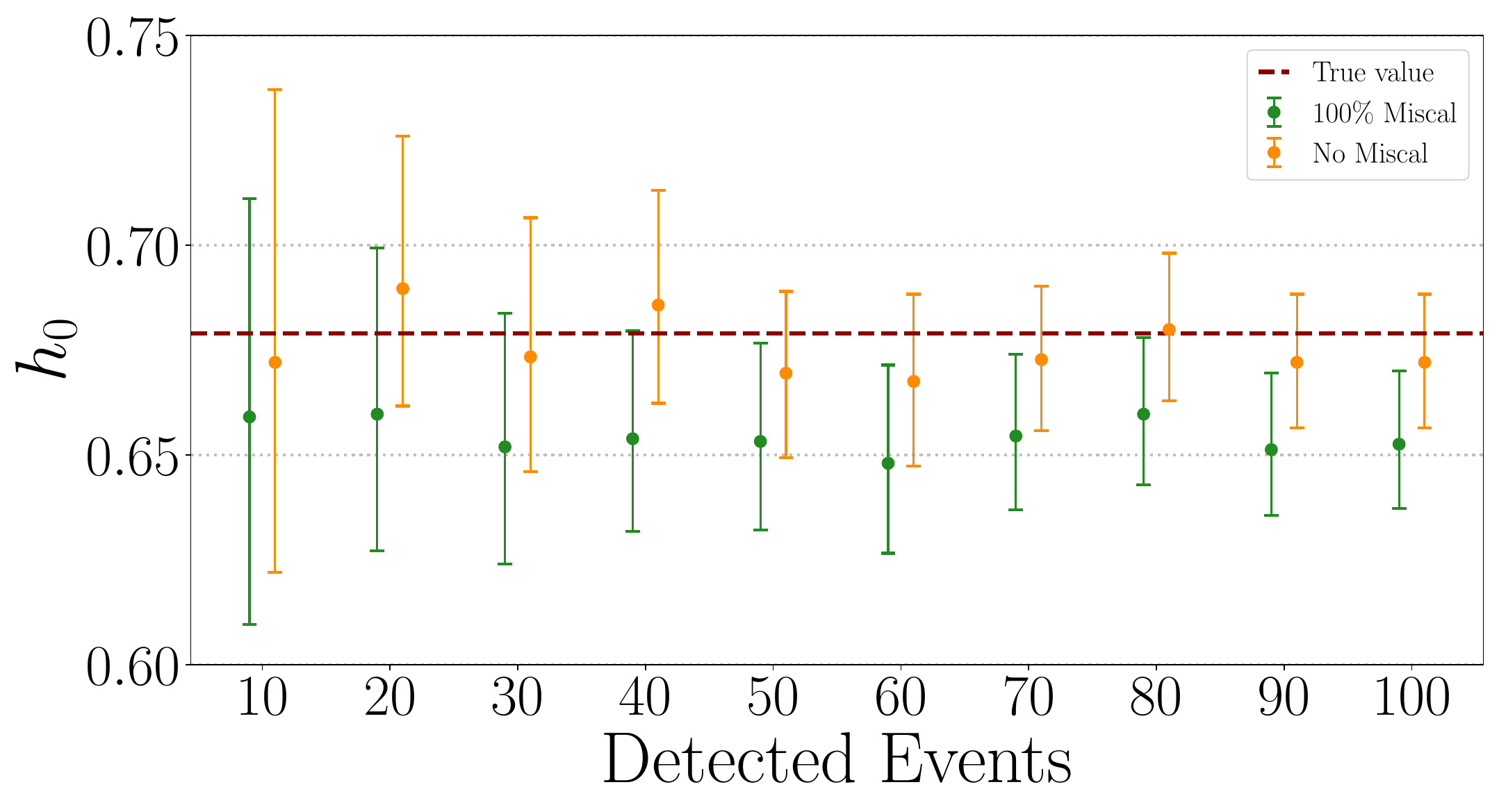}
   \vspace{-1.5\baselineskip}
   \caption{100\% of the events miscalibrated.}
   \label{Fig.h0_nEvents_cusRange_miscal1.0}  
\end{subfigure}
\caption{$h_0$ posteriors when we have a certain number of detections, plotted when 10\%, 50\% and 100\% of the events are miscalibrated. We introduce an artificial offset on the x axis for plotting purposes.}
\label{Fig.h0_nEvents_cusRange}
\end{figure}

\section{Conclusion}
GW observations of EM-bright compact binary coalescences provide an independent way to measure the Hubble constant and potentially break the existing tension between the early and late universe \hubble measurements. As we observe more of such events, the resulting \hubble posterior will be increasingly constraining, making it important to thoroughly control and understand potential systematic biases.

In this study, we applied large CEs to the GW data stream and investigated their effects on the inference of \hubble. %
Our analysis was constructed not to contain any systematic errors or statistical uncertainties from the EM observations like peculiar velocities~\cite{Howlett:2019mdh,Mukherjee:2019qmm} or viewing angles~\cite{Chen:2020dyt}, or from the GW measurements like waveform systematics, for our inference of \hubble. 

We found that the $H_0$ posterior does not exclude the true value from its 90\% CI, \emph{unless} we are inferring \hubble with more than 50 BNS detections that all suffer from the same large CEs, where the 90\% CI drops to below {4.4 $\mathrm{km \ s}^{-1}\mathrm{Mpc}^{-1}$}. The total number of detections needed for excluding the true $H_0$ from its 90\% CI increases to $\sim$ 90 BNSs when 50\% of them suffer from miscalibration.
When 10\% of BNSs are miscalibrated, the true value is not excluded even after 100 BNS detections. For comparison, systematic errors due to the EM observation, for example, kilonova viewing angles, will be 2\% after 50 BNS detections, although this bias may be mitigated by modeling the inclination distribution~\cite{Chen:2020dyt,chen2024mitigatingcounterpartselectioneffect}.

All of the outliers $\etaOut_i$ that motivated our study are based on estimated CE around times of real physical changes or malfunctions of the detectors during O3. These events are generally rare and relatively short lived;  typically $< 1\%$ of the time over any few-month observing period -- the typical duration of stable detector configurations during an observing run~\cite{Sun:2020wke, Sun:2021qcg}. In our analysis, we assume we only know and use $\etaTyp_i$ to marginalize over CEs during PE, while $\etaOut_i$, although present in some fraction of the detected BNS events, is assumed to be unknown and uncharacterized. Given the current estimate of astrophysical event rate, $320^{+490}_{-240}\;\mathrm{Gpc}^3 \mathrm{yr}^{-1}$~\cite{PhysRevX.11.021053}, there is a very low probability that a large CE remains uncharacterized over a period of stable configurations during which dozens of BNSs may be detected.

Our results imply that CEs are not going to be a significant concern in the measurement of the Hubble constant with the bright-siren method for the next many years. In the most realistic case, where large instances of CE like the ones described in this paper affect a small percent of the sources, CE will not become the limiting factor until more than 100 BNSs, each with an EM counterpart, have been found. Since the bright-siren method is likely to provide the smallest statistical uncertainties, other approaches to constrain \hubble using distance measurements from GW sources are going to be even less sensitive to CEs.

In this work, we have focused on the effect of CEs on inference results of BNSs. We will report on other types of compact binary coalescences, such as neutron star black hole mergers and binary black holes, as well as on calibration parameters from physiCal, in a forthcoming paper.

\textit{Software}. The authors acknowledge the use of the LIGO Algorithm Library~\cite{lalsuite}, and specifically of the \textsc{LALInference} inference package~\cite{Veitch:2014wba}, as released through \textsc{Anaconda}~\cite{anaconda}. 
Plots were produced with \textsc{matplotlib}~\cite{Hunter:2007}. 
The authors acknowledge use of \textsc{iPython}~\cite{PER-GRA:2007}, \textsc{NumPy}~\cite{harris2020array} and \textsc{SciPy}~\cite{Virtanen:2019joe}.

\section{Acknowledgments}
The authors would like to thank Jonathan Gair, Evan Goetz, Martin Hendry, Daniel Holz, Suvodip Mukherjee, Ethan Payne, Jameson Graef Rollins, Nicola Tamanini, Paxton Turner, Madeline Wade, and Alan J. Weinstein for useful discussions. 

The authors are grateful for computational resources provided by the LIGO Laboratory and supported by National Science Foundation Grants PHY-0757058 and PHY-0823459. Y.~H., C.-~J.~H., S.~V.~, and J.~S.~K.~acknowledge support of the National Science Foundation, and the LIGO Laboratory. 
LIGO was constructed by the California Institute of Technology and Massachusetts Institute of Technology with funding from the National Science Foundation and operates under Cooperative Agreement No. PHY-2309200. Virgo is funded by the French Centre National de Recherche Scientifique (CNRS), the Italian Istituto Nazionale della Fisica Nucleare (INFN), and the Dutch Nikhef, with contributions by Polish and Hungarian institutes.
S.~V.and Y.~H.~acknowledge support of the National Science Foundation through the Award No. PHY-2045740.
C.-~J.~H. acknowledges the support from the Nevada Center for Astrophysics, from NASA Grant No. 80NSSC23M0104, and the National Science Foundation through the Award No. PHY-2409727.
H.-Y.~C. was supported by NASA through NASA Hubble Fellowship grants No.\ HST-HF2-51452.001-A awarded by the Space Telescope Science Institute, which is operated by the Association of Universities for Research in Astronomy, Inc., for NASA, under Contract No. NAS5-26555. H.-Y.~C is supported by the National Science Foundation under Grant No. PHY-2308752 and Department of Energy under Grant No.~DE-SC0025296.
L. S. acknowledges the support of the Australian Research Council Centre of Excellence for Gravitational Wave Discovery (OzGrav), Project No. CE170100004 and No. CE230100016, and the Australian Research Council Discovery Early Career Researcher Award, Project No. DE240100206.
This is LIGO Document No. DCC-P2100350.

\appendix
\clearpage
\null
\section{\miscal, \etaOut, and \etaTyp for the other five realizations}
\label{App.rdis}

Here, in Figs. \ref{Fig.all_mis_1238047284_H1}--\ref{Fig.all_mis_1257808500_H1}, we show the \miscal, \etaOut, and \etaTyp for the other five outlier cases identified for CEs during O3.

\begin{figure}[!htbp]
\centering
\begin{subfigure}{0.45\textwidth}
   \includegraphics[width=1\linewidth]{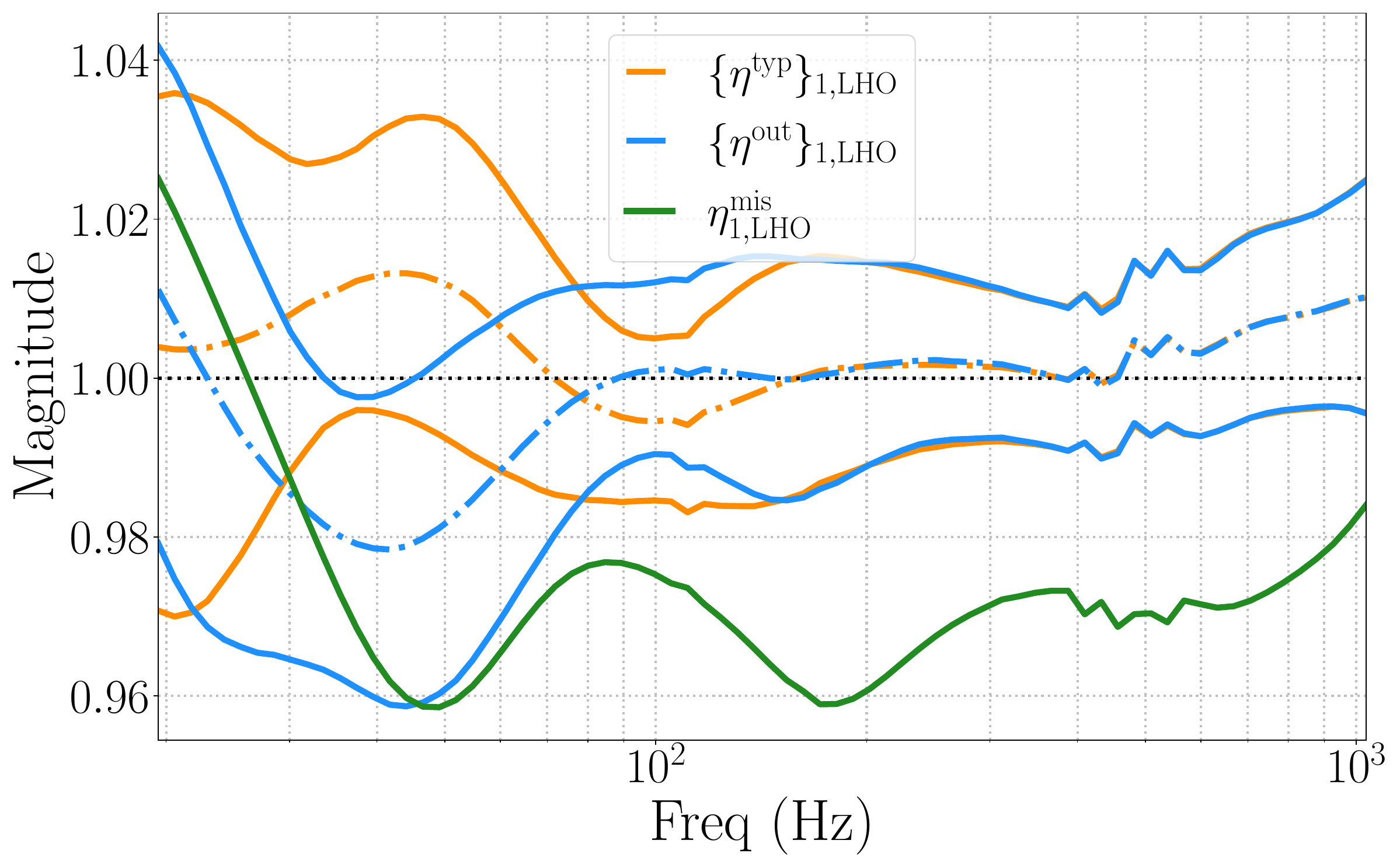}
   \vspace{-1.5\baselineskip}
   \caption{Amplitude.}
   \label{Fig.all_mis_amp_1238047284_H1} 
\end{subfigure}
\begin{subfigure}{0.45\textwidth}
   \includegraphics[width=1\linewidth]{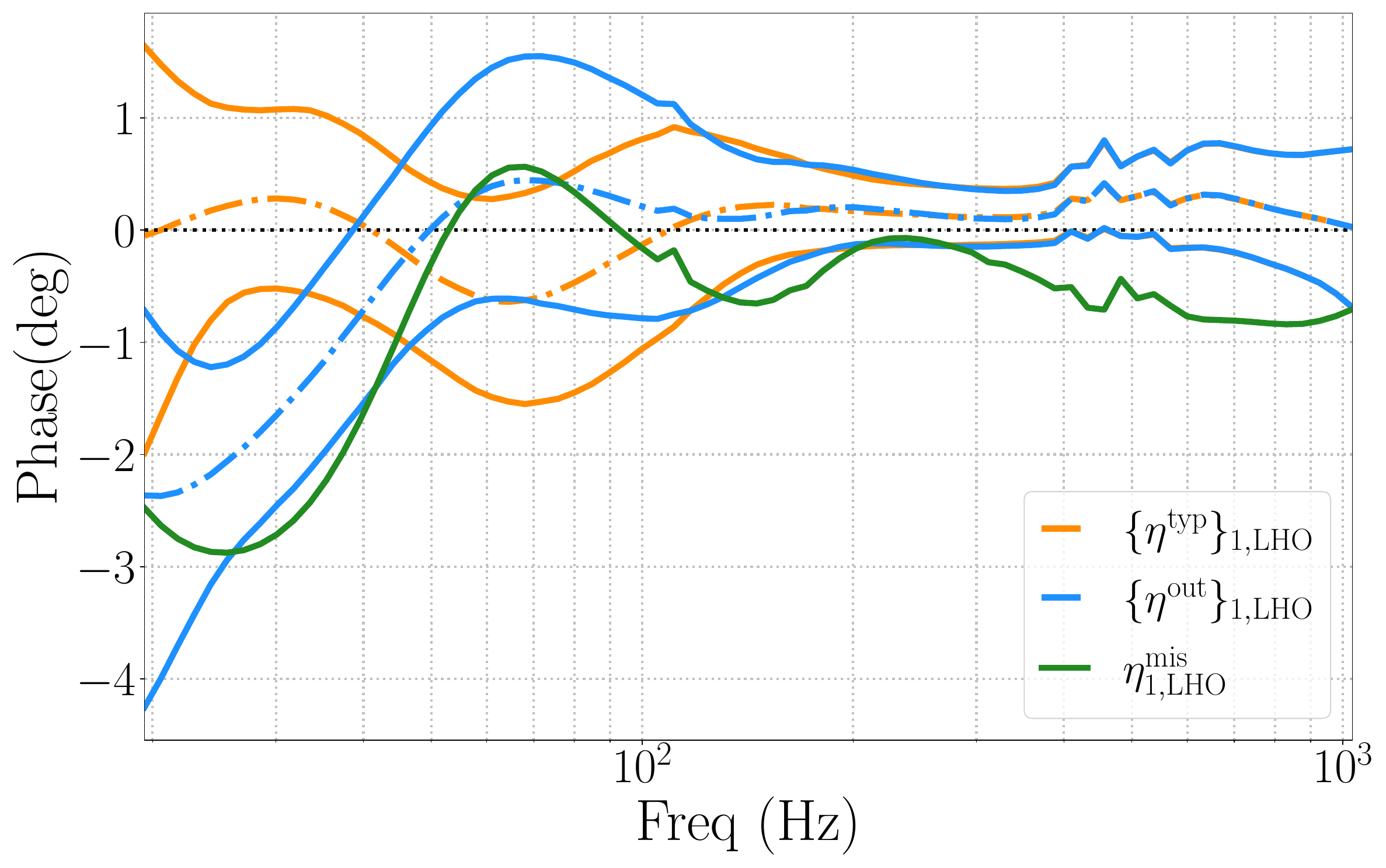}
   \vspace{-1.5\baselineskip}
   \caption{Phase.}
   \label{Fig.all_mis_pha_1238047284_H1}  
\end{subfigure}
\caption{Large CEs, $\etaOut_\mathrm{1,LHO}$(blue), compared to the corresponding typical distribution, $\etaTyp_\mathrm{1,LHO}$ (orange), both showing the edges of the 1-$\sigma$ CIs in each frequency bin. $\miscal_\mathrm{1,LHO}$ is plotted in green. }
\label{Fig.all_mis_1238047284_H1}
\end{figure}

\begin{figure}[!htbp]
\centering
\begin{subfigure}[b]{0.45\textwidth}
   \includegraphics[width=1\linewidth]{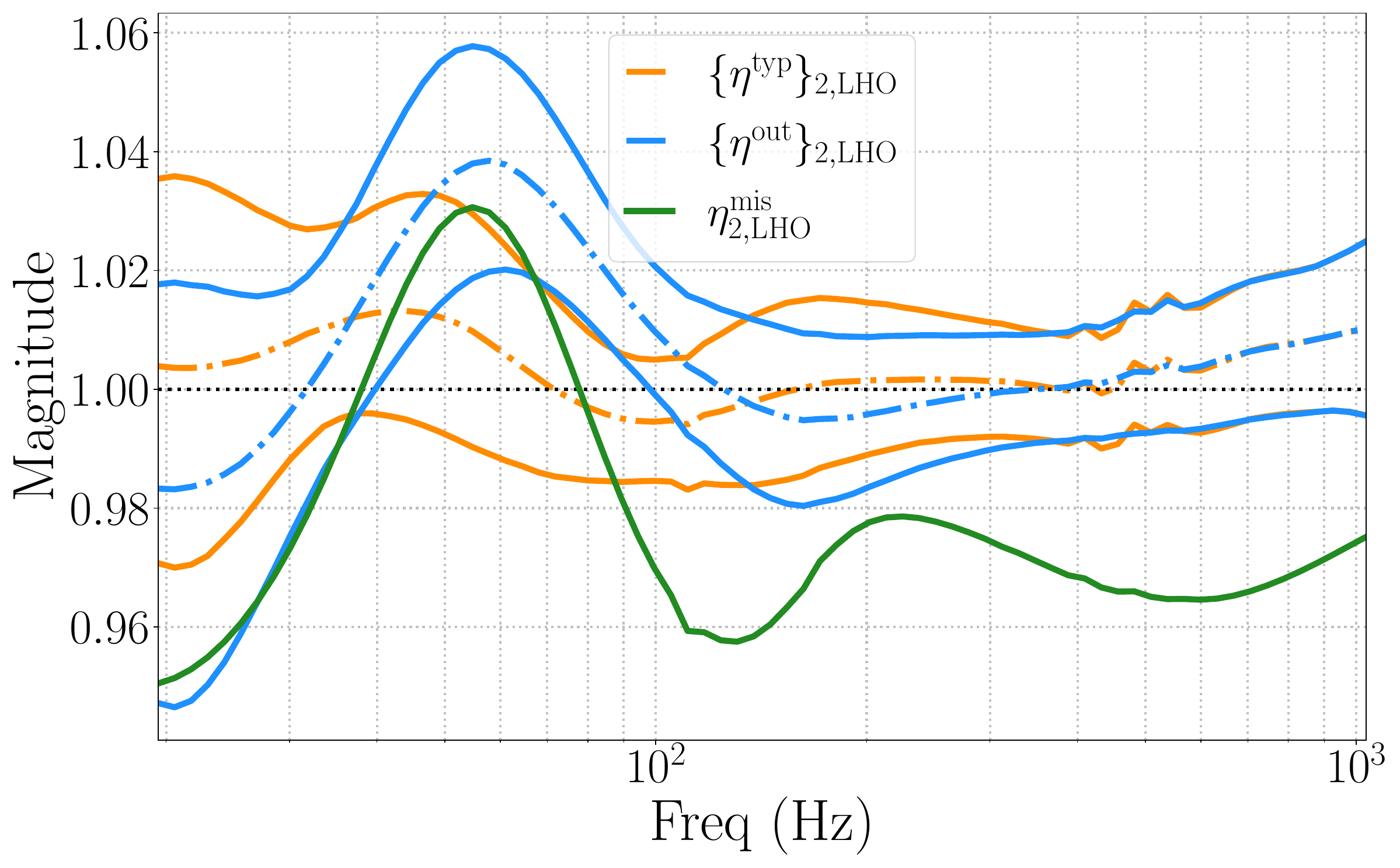}
   \vspace{-1.5\baselineskip}
   \caption{Amplitude.}
   \label{Fig.all_mis_amp_1242116340_H1.} 
\end{subfigure}
\begin{subfigure}[b]{0.45\textwidth}
   \includegraphics[width=1\linewidth]{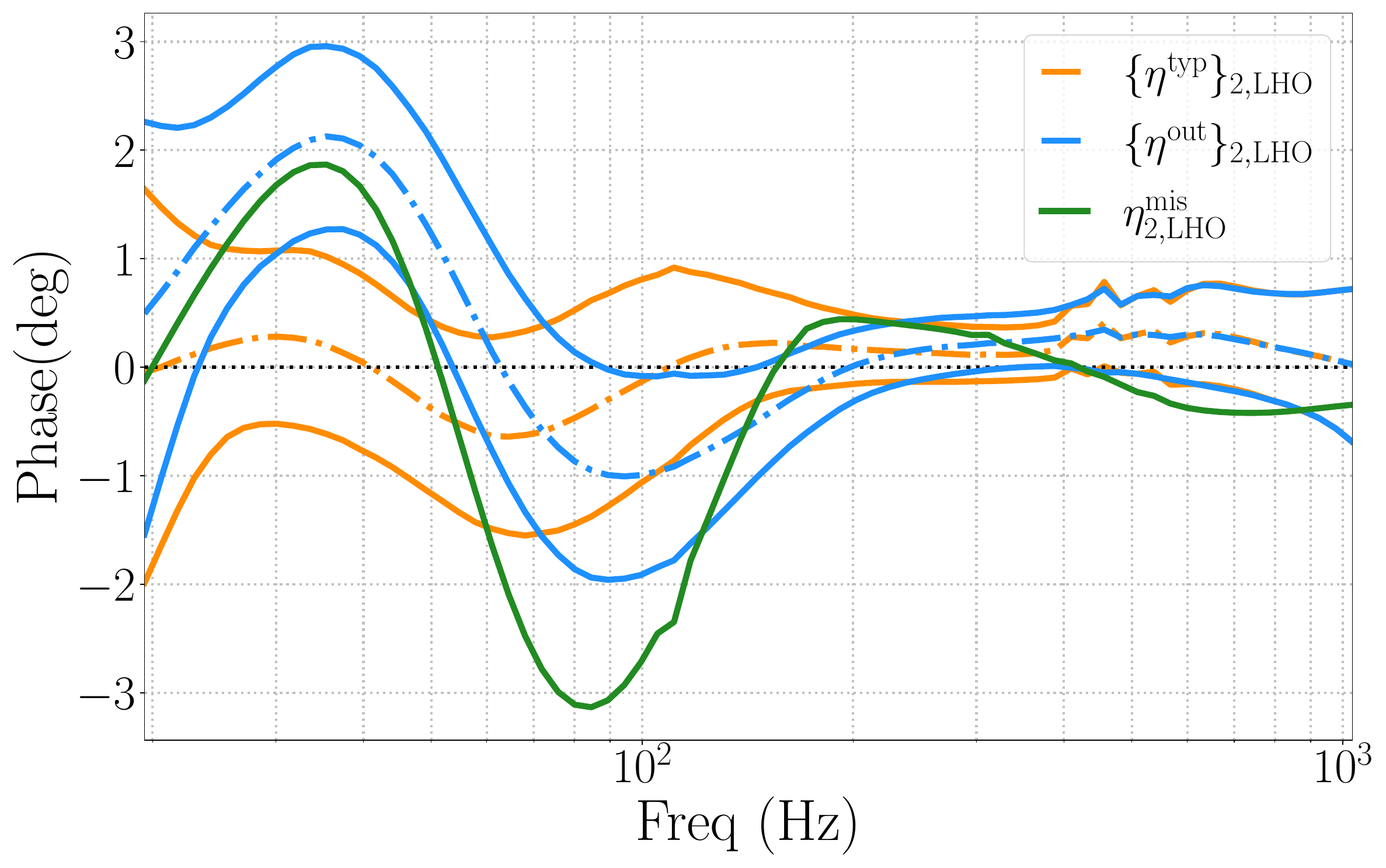}
   \vspace{-1.5\baselineskip}
   \caption{Phase.}
   \label{Fig.all_mis_pha_1242116340_H1}  
\end{subfigure}
\caption{Large CEs, $\etaOut_\mathrm{2,LHO}$(blue), compared to the corresponding typical distribution, $\etaTyp_\mathrm{2,LHO}$ (orange), both showing the edges of the 1-$\sigma$ CIs in each frequency bin. $\miscal_\mathrm{2,LHO}$ is plotted in green.}
\label{Fig.all_mis_1242116340_H1}
\end{figure}

\begin{figure}[!htbp]
\centering
\begin{subfigure}[b]{0.45\textwidth}
   \includegraphics[width=1\linewidth]{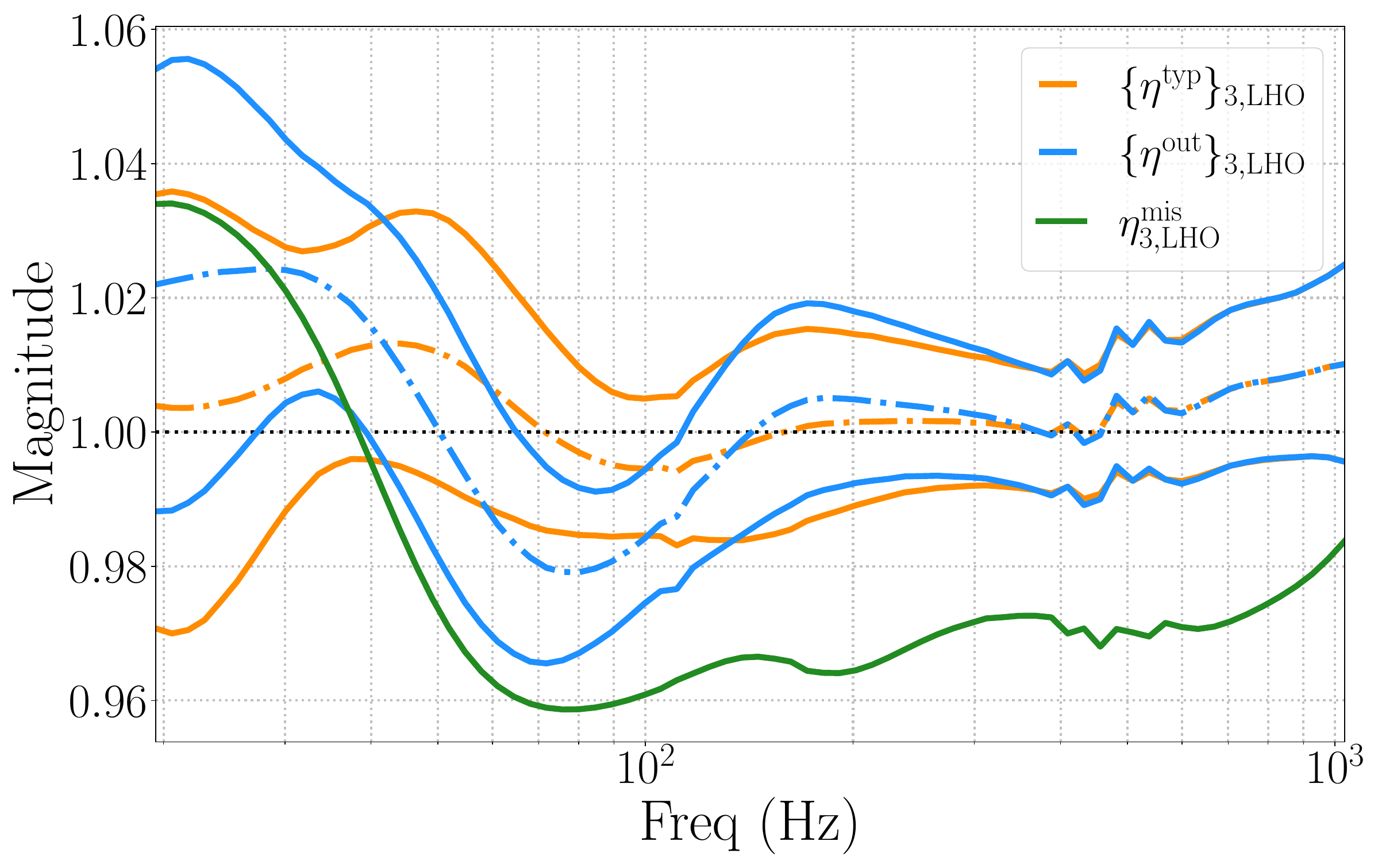}
   \vspace{-1.5\baselineskip}
   \caption{Amplitude.}
   \label{Fig.all_mis_amp_1242701640_H1} 
\end{subfigure}
\begin{subfigure}[b]{0.45\textwidth}
   \includegraphics[width=1\linewidth]{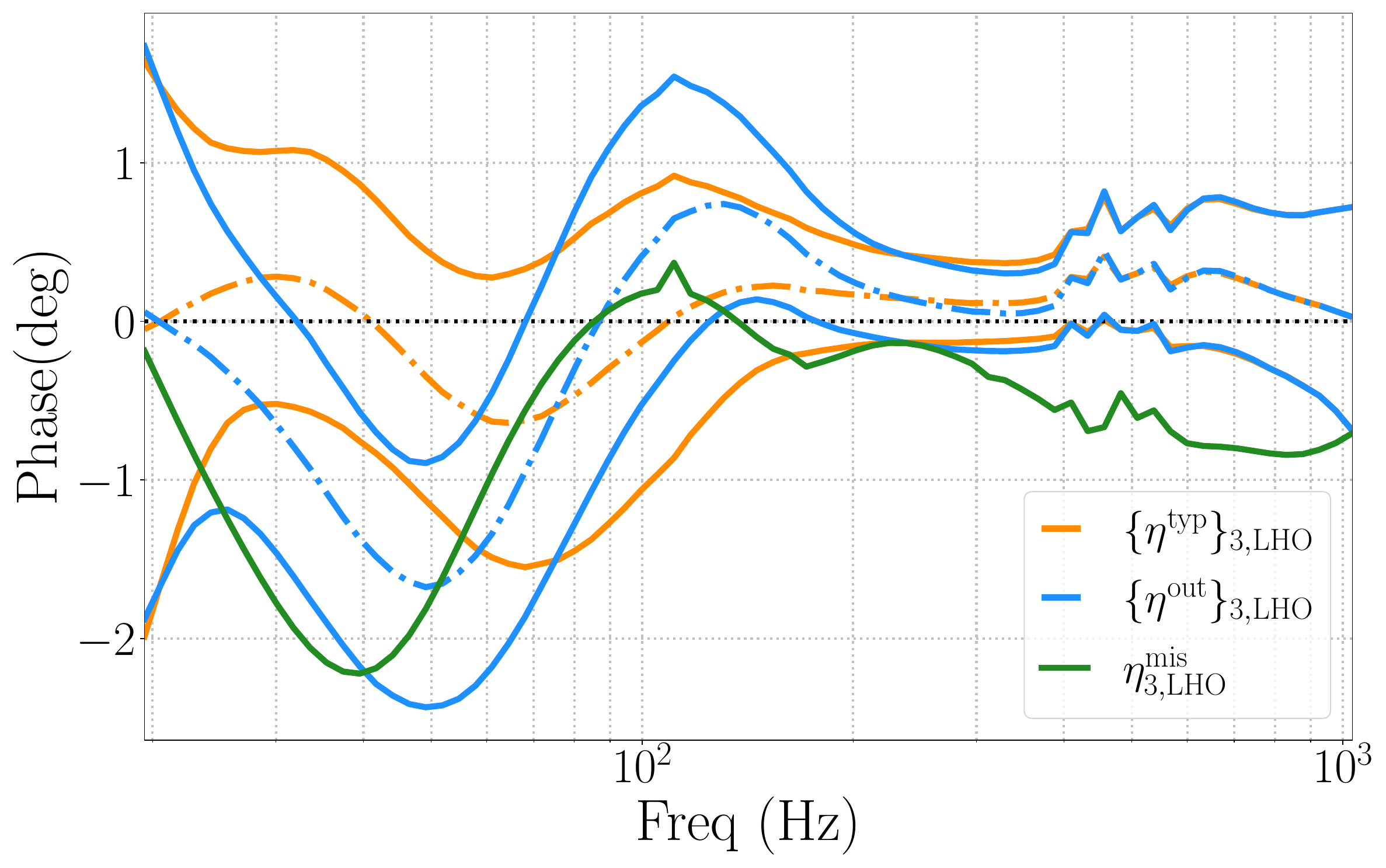}
   \vspace{-1.5\baselineskip}
   \caption{Phase.}
   \label{Fig.all_mis_pha_1242701640_H1}  
\end{subfigure}
\caption{Large CEs, $\etaOut_\mathrm{3,LHO}$(blue), compared to the corresponding typical distribution, $\etaTyp_\mathrm{3,LHO}$ (orange), both showing the edges of the 1-$\sigma$ CIs in each frequency bin. $\miscal_\mathrm{3,LHO}$ is plotted in green.}\label{Fig.all_mis_1242701640_H1}
\end{figure}

\begin{figure}[!htbp]
\centering
\begin{subfigure}[b]{0.45\textwidth}
   \includegraphics[width=1\linewidth]{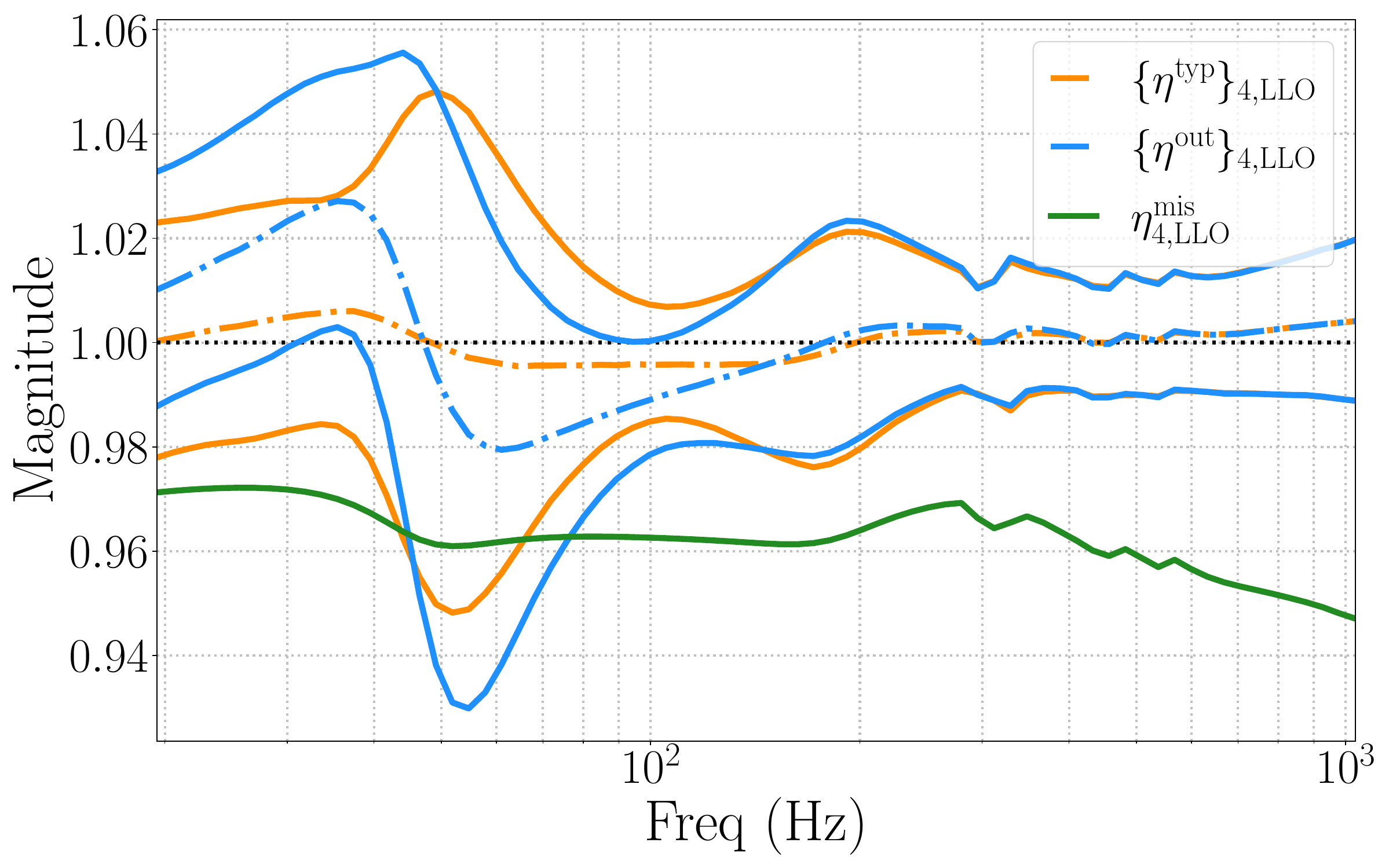}
   \vspace{-1.5\baselineskip}
   \caption{Amplitude.}
   \label{Fig.all_mis_amp_1243790460_L1} 
\end{subfigure}
\begin{subfigure}[b]{0.45\textwidth}
   \includegraphics[width=1\linewidth]{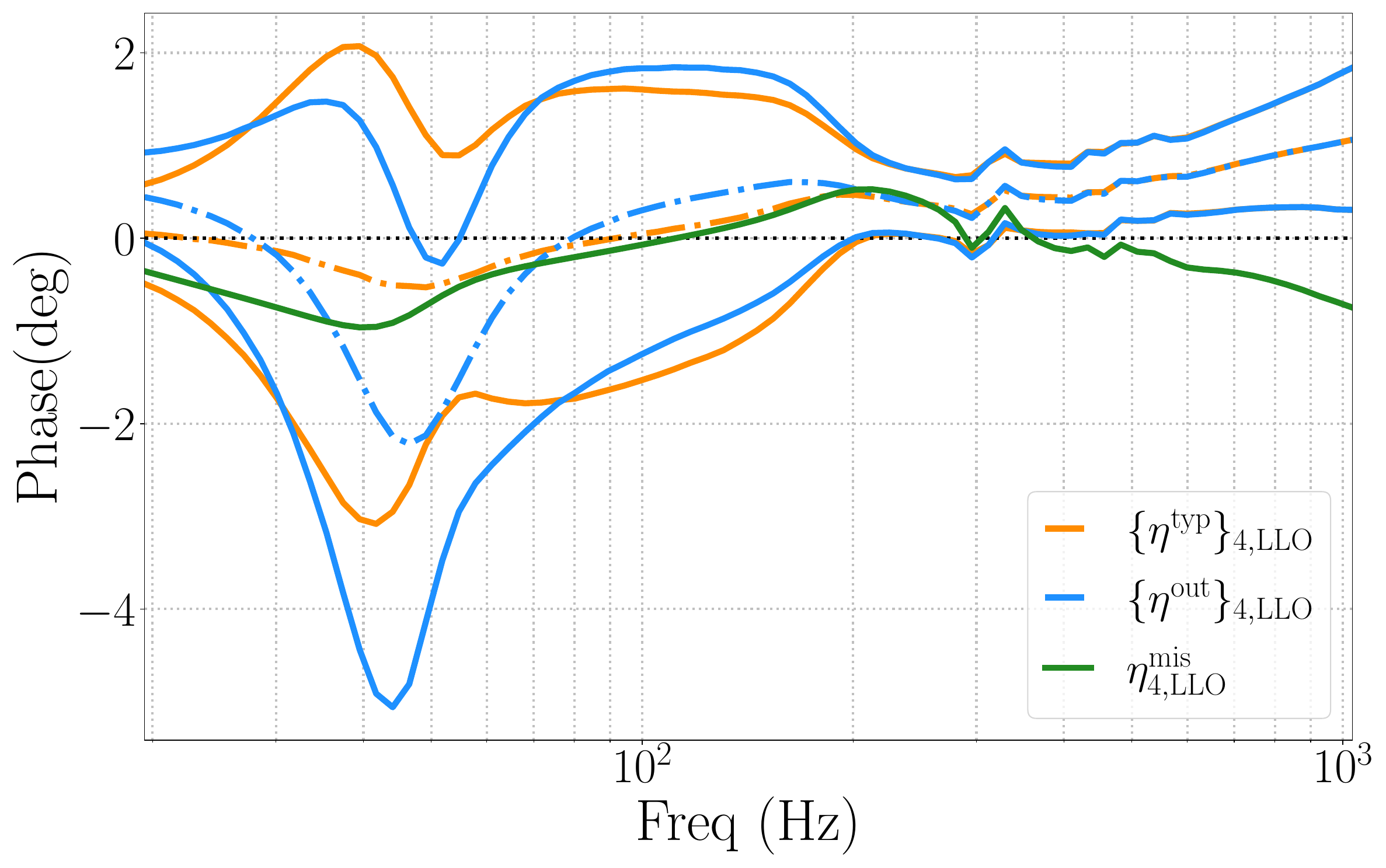}
   \vspace{-1.5\baselineskip}
   \caption{Phase.}
   \label{Fig.all_mis_pha_1243790460_L1}  
\end{subfigure}
\caption{Large CEs, $\etaOut_\mathrm{4,LLO}$(blue), compared to the corresponding typical distribution, $\etaTyp_\mathrm{4,LLO}$ (orange), both showing the edges of the 1-$\sigma$ CIs in each frequency bin. $\miscal_\mathrm{4,LLO}$ is plotted in green.}\label{Fig.all_mis_1243790460_L1}
\end{figure}

\begin{figure}[!htbp]
\centering
\begin{subfigure}[b]{0.45\textwidth}
   \includegraphics[width=1\linewidth]{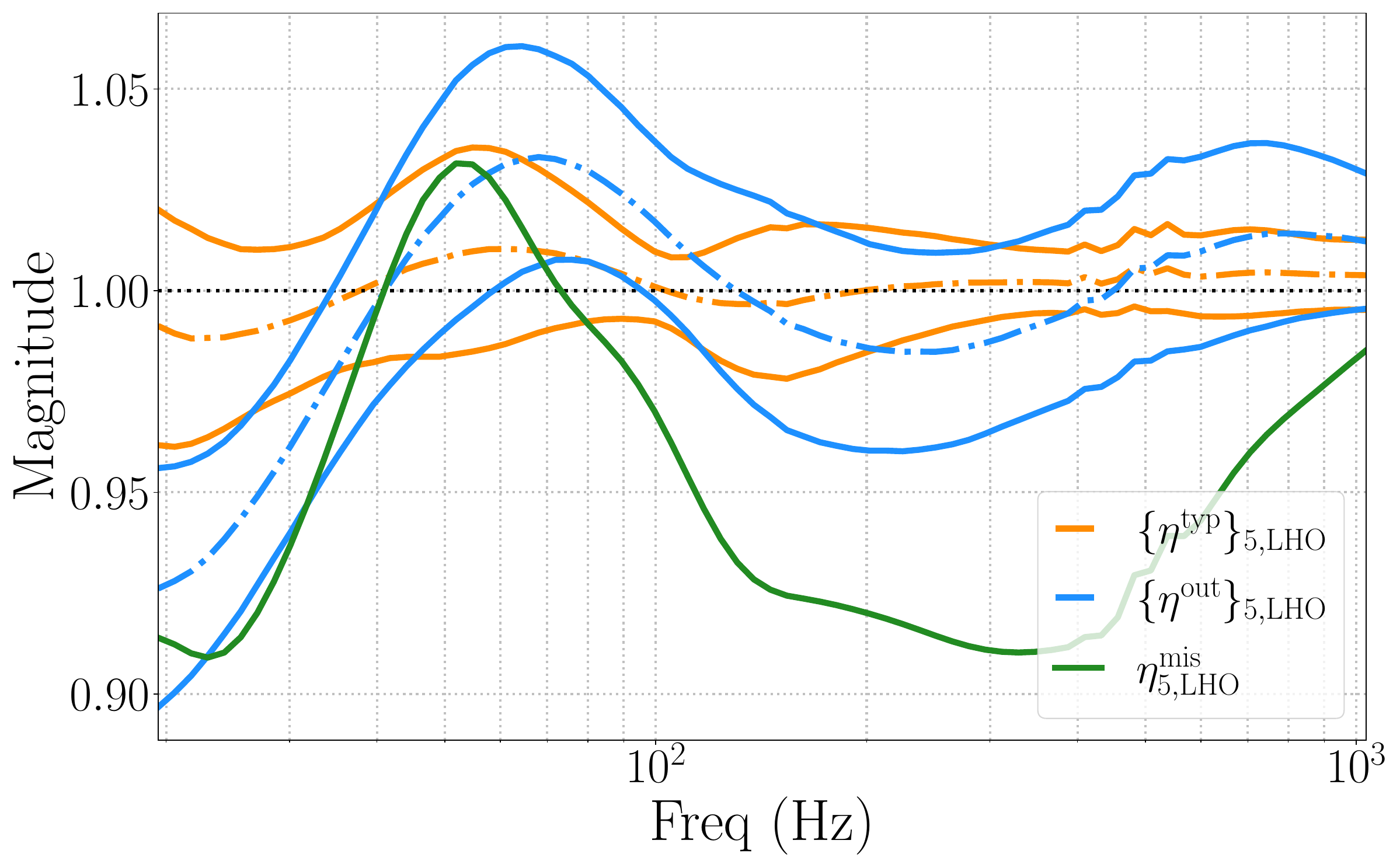}
   \vspace{-1.5\baselineskip}
   \caption{Amplitude.}
   \label{Fig.all_mis_amp_1257808500_H1} 
\end{subfigure}
\begin{subfigure}[b]{0.45\textwidth}
   \includegraphics[width=1\linewidth]{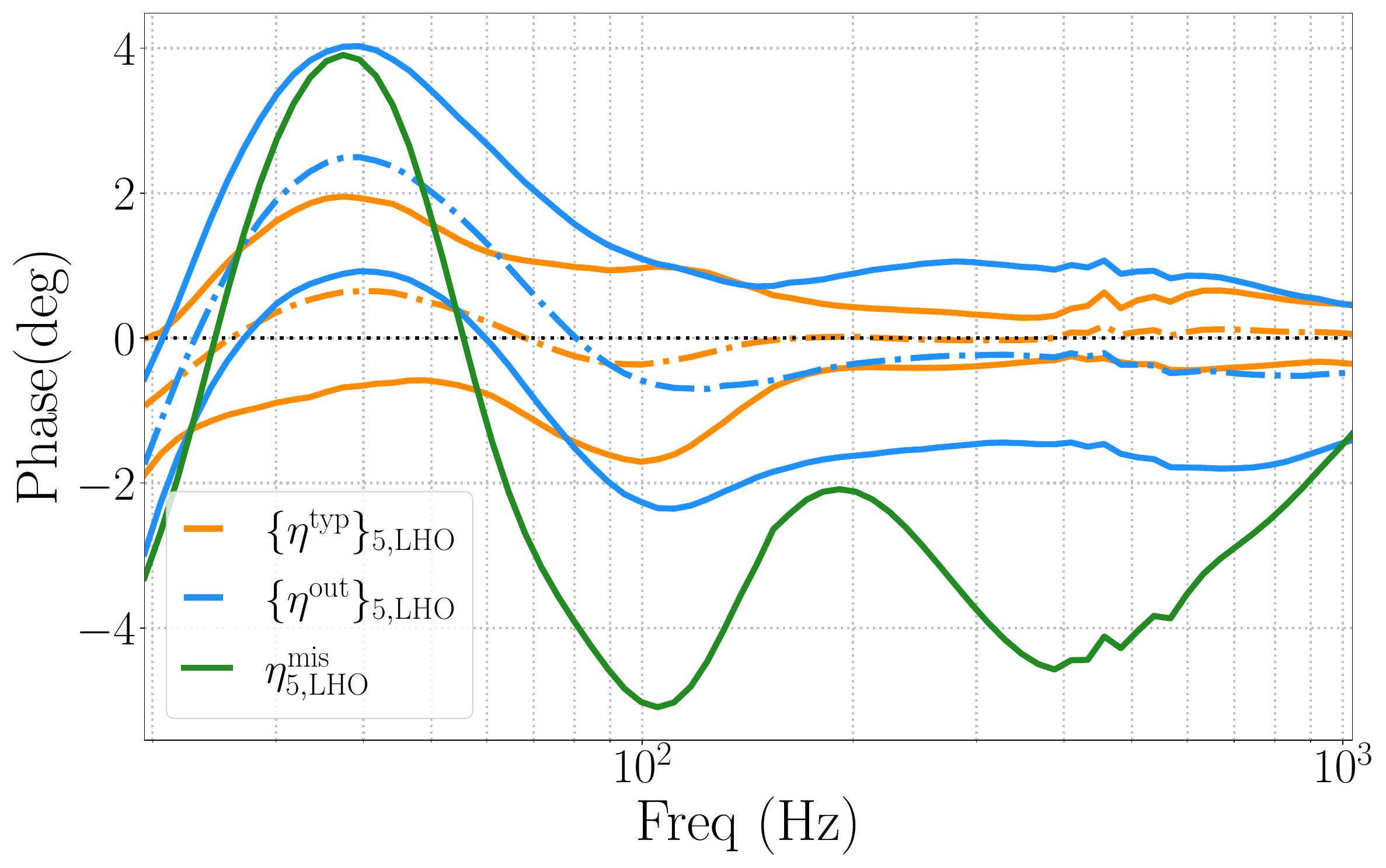}
   \vspace{-1.5\baselineskip}
   \caption{Phase.}
   \label{Fig.all_mis_pha_1257808500_H1}  
\end{subfigure}
\caption{Large CEs, $\etaOut_\mathrm{5,LHO}$(blue), compared to the corresponding typical distribution, $\etaTyp_\mathrm{5,LHO}$ (orange), both showing the edges of the 1-$\sigma$ CIs in each frequency bin. $\miscal_\mathrm{5,LHO}$ is plotted in green.}
\label{Fig.all_mis_1257808500_H1}
\end{figure}

\section{Spline Results}
\label{App.Spline}

In Fig.~\ref{Fig.dist_BNS_SNR50_Spline}, we show the distance likelihoods for BNSs at an SNR of 50, where the green and purple shaded distributions are obtained from the runs with the physiCal and Spline methods, respectively, both miscalibrated by the same \miscal. We report $\Delta \dl$ in Table~\ref{Table.dl bias Spline}. The differences between the results are quite small compared to those between the physiCal runs with and without CEs.

\begin{figure}
\centering
\includegraphics[width=\linewidth]{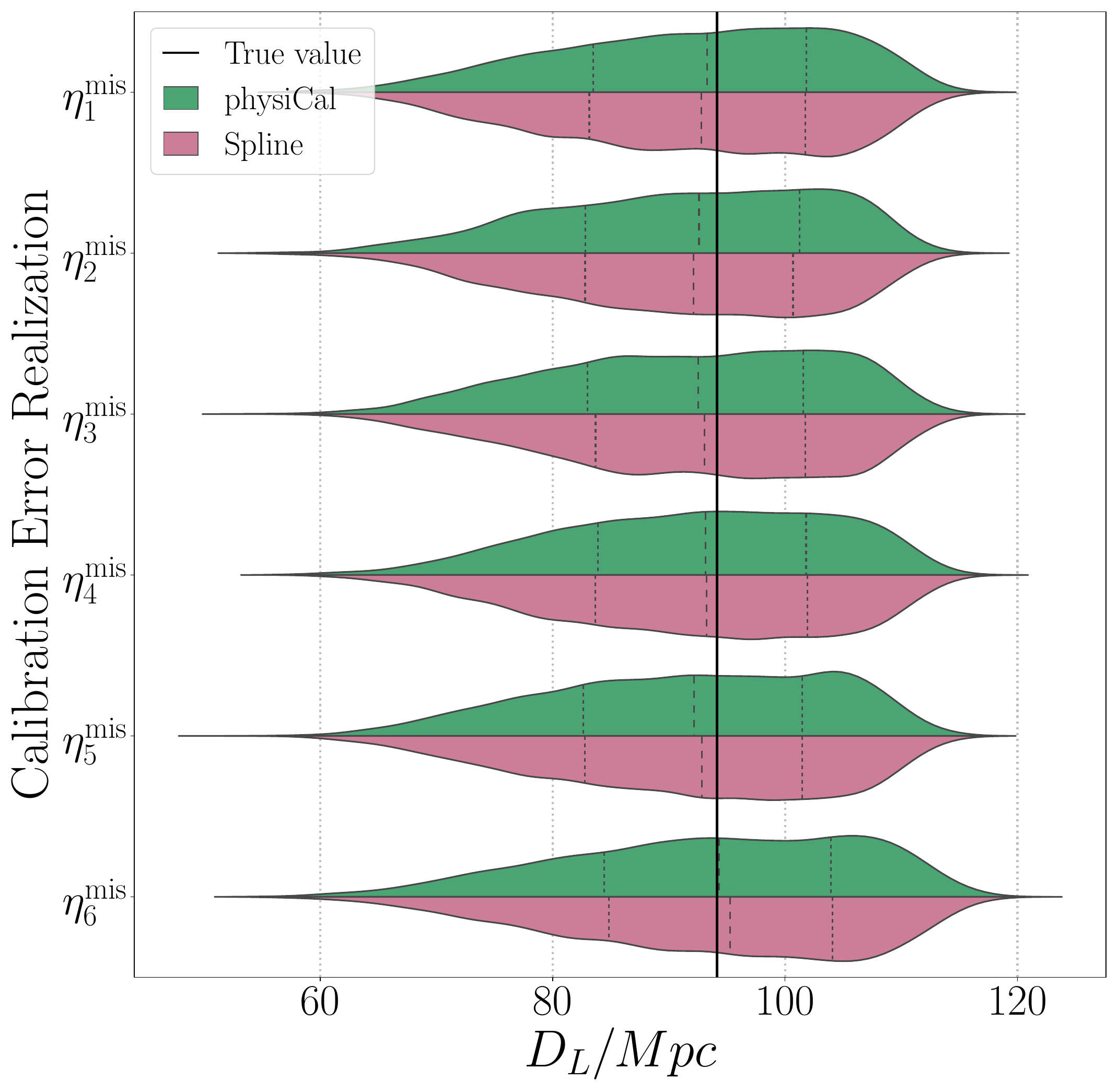}
\caption{\dl likelihoods for the six scenarios, miscalibrated, physiCal (green) vs. Spline (purple) runs, the vertical dashed lines mark the 25\%, 50\% and 75\% percentiles.}
\label{Fig.dist_BNS_SNR50_Spline}
\end{figure}

\begin{table}[t]
\centering
\begin{tabularx}{0.90\linewidth}{b|bb}
\toprule\toprule
CE realization & physiCal & Spline \\
\midrule
$\miscal_1$&$-0.8\%$&$-1.5\%$\\
\midrule
$\miscal_2$&$-2.0\%$&$-2.6\%$\\
\midrule
$\miscal_3$&$-1.6\%$&$-1.4\%$\\
\midrule
$\miscal_4$&$-0.8$\%&$-0.6\%$\\
\midrule
$\miscal_5$&$-1.7\%$&$-1.2\%$\\
\midrule
$\miscal_6$&$0.2\%$&$1.0\%$\\
\bottomrule\bottomrule
\end{tabularx}
\caption{$\Delta \dl$ in the likelihoods for physiCal vs. Spline results. }
\label{Table.dl bias Spline}
\end{table}

\bibliographystyle{apsrev4-1}
\bibliography{ThisPaper}

\end{document}